\documentclass[10pt,journal,compsoc]{IEEEtran}

\usepackage[utf8]{inputenc}
\usepackage{multirow}
\usepackage{tabularx}
\usepackage[table, dvipsnames]{xcolor}
\usepackage{pdflscape}
\usepackage[innerleftmargin=\parindent]{mdframed}
\usepackage{enumitem}
\usepackage[hidelinks]{hyperref}
\usepackage{graphicx}
\usepackage{subcaption}
\usepackage{amssymb}
\usepackage{amsmath}
\usepackage{cleveref}
\usepackage{comment}
\usepackage{balance}

\crefformat{section}{\S#2#1#3} % see manual of cleveref, section 8.2.1
\crefformat{subsection}{\S#2#1#3}
\crefformat{subsubsection}{\S#2#1#3}

\usepackage[ruled,linesnumbered]{algorithm2e}
\SetKw{KwFrom}{from}
\SetKw{KwUpdate}{ClientUpdate}
\SetKw{KwLineBreak}{\\}

\definecolor{grad_0}{RGB}{235, 130, 128} %{100,245,100}
\definecolor{grad_1}{RGB}{240, 185, 147} %{125,245,125}
\definecolor{grad_2}{RGB}{245, 226, 150} %{150,250,150}
\definecolor{grad_3}{RGB}{225, 235, 185} %{175,250,175}
\definecolor{grad_4}{RGB}{220, 245, 215} %{200,255,200}

\newmdenv[topline=false,
  innerleftmargin=1em,
  bottomline=false, rightline=false,
  skipabove=0.5em, skipbelow=1em]{leftlineframed}

\title{The Disruptions of 5G on \\ Data-driven Technologies and Applications}
\author{ Dumitrel~Loghin,~\IEEEmembership{Member,~IEEE},
Shaofeng~Cai, Gang~Chen,~\IEEEmembership{Member,~IEEE}, Tien~Tuan~Anh~Dinh, Feiyi~Fan,
Qian~Lin, Janice~Ng, Beng~Chin~Ooi,~\IEEEmembership{Fellow,~IEEE}, Xutao~Sun,
Quang-Trung~Ta, Wei~Wang, Xiaokui~Xiao, Yang~Yang,
Meihui~Zhang~\IEEEmembership{Member,~IEEE}, Zhonghua~Zhang\\
\IEEEcompsocitemizethanks{\IEEEcompsocthanksitem D. Loghin, S. Cai, F. Fan, Q.
Lin, B.C. Ooi, X. Sun, Q.-T. Ta, W. Wang, X. Xiao, and Z. Zhang are with
National University of Singapore, Singapore 117417.
E-mail: [dumitrel, shaofeng, fanfy, linqian, ooibc, sunxt, taqt, wangwei,
xkxiao]@comp.nus.edu.sg; idmzhz@nus.edu.sg \IEEEcompsocthanksitem G. Chen is
with Zhejiang University, Hangzhou 310027, China. E-mail:
cg@cs.zju.edu.cn \IEEEcompsocthanksitem T.T.A. Dinh is with Singapore University
of Technology \& Design, Singapore 487372. Email: dinhtta@sutd.edu.sg
\IEEEcompsocthanksitem J. Ng is with UC Berkeley, USA, and was an intern in
School of Computing, National University of Singapore, during this project.
Email: janiceng@berkeley.edu \IEEEcompsocthanksitem Y. Yang is with University
of Electronic Science and Technology, Chengdu, Sichuan, China 611731.
Email: dlyyang@gmail.com \IEEEcompsocthanksitem M. Zhang is with Beijing
Institute of Technology, Beijing, China 100081.
Email: meihui\_zhang@bit.edu.cn }
\thanks{Manuscript received September 19, 2019.}
}

\begin{document}

\maketitle

\begin{abstract}

With 5G on the verge of being adopted as the next mobile network, there is a
need to analyze its impact on the landscape of computing and data management. In
this paper, we analyze the impact of 5G on both traditional and emerging
technologies and project our view on future research challenges and
opportunities. With a predicted increase of 10-100x in bandwidth and 5-10x
decrease in latency, 5G is expected to be the main enabler for smart cities,
smart IoT and efficient healthcare, where machine learning is conducted at the
edge. In this context, we investigate how 5G can help the development of
federated learning. Network slicing, another key feature of 5G, allows running
multiple isolated networks on the same physical infrastructure. However,
security remains the main concern in the context of virtualization,
multi-tenancy and high device density. Formal verification of 5G networks can be
applied to detect security issues in massive virtualized environments. In
summary, 5G will make the world even more densely and closely connected. What we
have experienced in 4G connectivity will pale in comparison to the vast amounts
of possibilities engendered by 5G.

\end{abstract}

\begin{IEEEkeywords}
5G mobile communication, database systems, network slicing, Internet of
Things, edge computing, federated learning, data privacy, security management.
\end{IEEEkeywords}

\IEEEpeerreviewmaketitle

\section{Introduction}

% Motivation (impact on economy/life style/ etc)
Fifth Generation (5G) mobile communication technologies are on the way to be
adopted all over the world. At the moment, 5G is being deployed in small areas
in almost all the continents, with a higher number of available networks in
Europe and the USA \cite{5g_map}. In future, 5G is predicted to account for at
least 15\% of the total mobile communications market by 2025
\cite{5g_market_forbes}. It is therefore timely to analyze the impact of 5G on
key areas of research related to data management and processing, including
databases, distributed systems, blockchain, and machine learning.

% Opportunities
With its increased bandwidth of up to 20 Gigabits per second (Gbps), low latency
of 1 millisecond (ms), high device density of one million devices per square
kilometer, and virtualization technologies \cite{min_int2020}, 5G is generating
new opportunities in computing. New use cases, such as remote healthcare based
on virtual reality (VR) and augmented reality (AR), or ultra-high-definition
(UHD) movie streaming can only be possible in 5G networks \cite{5g_ch_opp}.
Other applications, such as machine-to-machine (M2M) communication in automotive
and smart drones, and high-density Internet of Things (IoT) devices in smart
cities can be handled by the current technologies, such as 4G, WiFi, and
Bluetooth, but they can greatly benefit from the improvements of 5G
\cite{5g_ch_opp}.

% Contribution: measurements
In this paper, we provide performance measurements done in a real 5G network
showing a maximum download bandwidth of 458 Megabits per second (Mbps) and
minimum round-trip time (RTT) of 6 ms. While these numbers are still far from
the 5G specifications \cite{min_int2020}, they represent current 5G networks
running in Non-Standalone (NSA) mode and expose more than $5\times$ better
performance, in terms of bandwidth and latency, compared to 4G networks. We plan
to use these performance measurements to emulate 5G deployments where data
management and processing systems can be evaluated.

% trend/projection/vision
Beyond the obvious impact of 5G in well-known areas, we examine the
opportunities and challenges in computing areas related to distributed data
management and processing. For example, 5G technology has the potential to bring
forth the idea of millions of shared (micro-)databases which will impact data
analytics, federated learning \cite{federated_learning_19}, and security at the
edge. Nonetheless, the concept of millions of databases poses challenges in
terms of privacy and security.

% Contribution and Challenges #1 - Security
In this paper, we conduct a systematic survey of challenges and opportunities 5G
is bringing to key areas in computing, such as edge computing and IoT
(\cref{sec:edge}), networking (\cref{sec:netverif}), data storage and processing
(\cref{sec:db}), blockchain (\cref{sec:blockchain}), artificial intelligence
(\cref{sec:ai}), and security and privacy (\cref{sec:security}). We highlight
security as a major challenge in 5G deployments, due to multiple factors. First,
the high density and large number of IoT devices that can be connected to a 5G
network will increase the risk of attacks, such as Distributed Denial-of-Service
(DDoS). It is well-known that IoT devices are easier to break and that some of
the largest-scale attacks were conducted using distributed IoT
devices~\cite{ddos_botnet}. Second, the full virtualization in 5G networks is
posing new challenges in security management. We analyze in this paper what are
the risks of network slicing \cite{nslicing_survey}, the key technology in 5G
virtualization.

% Challenge #2 - backhaul and backbone connectivity
Another key challenge is the current backhaul and cloud network infrastructure
that is not able to cope with the increased traffic generated by the 5G mobile
networks. Our measurements of inter-region cloud connections show that the
throughput is almost always lower than 100 Mbps, while the latency exceeds 300
ms in some cases. These values are far behind the requirements of 5G. In a study
by McKinsey, it is estimated that an operator needs to spend up to 300\% more on
infrastructure to cope with a 50\% increase in data volume \cite{5ginfra}. This,
together with virtualization, introduces new challenges in the delivery and
monetization of 5G services, while delaying the adoption of 5G.

% Organization
The remainder of this paper is organized as follows. In Section~\ref{sec:5g}, we
review 5G specifications, evaluate current deployments with performance
measurements and examine 5G network simulators. In Section~\ref{sec:impact}, we
analyze the impact of 5G on major computing areas that are related to data
processing and management, such as edge computing and IoT, network verification,
databases, blockchain, federated learning, security and privacy. We end
Section~\ref{sec:impact} by discussing challenges and opportunities. In
Section~\ref{sec:use_cases}, we discuss how 5G is going to boost new use cases,
such as telemedicine, AR/VR, e-commerce, fintech, smart cars, smart drones, and
smart cities. We conclude the paper in Section~\ref{sec:conclusion}.

\section{5G Technologies}
\label{sec:5g}

In this section, we review the properties of 5G, in comparison with the previous
generations of mobile and wireless technologies. We provide a summary of current
5G deployments, with performance measurements done with two 5G smartphones, and
explore solutions for simulating 5G networks.

\subsection{An Overview of 5G}
\label{sec:5g_specs}
% Xutao & Dumi

\begin{figure}[tp]
\centering
\includegraphics[width=0.485\textwidth]{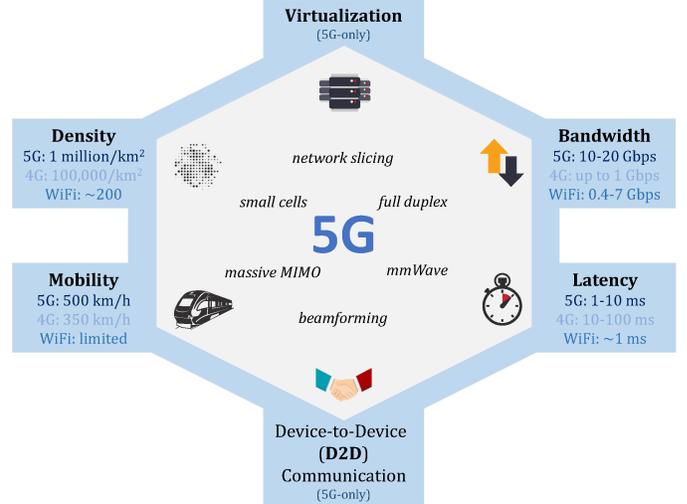}
\caption{5G overview}
\label{fig:overview}
\end{figure}

5G is the fifth generation of cellular network technologies specified by the 3rd
Generation Partnership Project (3GPP). It proceeds 2G, 3G, and 4G and their
associated technologies, while introducing significant performance improvements,
as shown in \autoref{fig:overview} and \autoref{tab:mobile_comm_comp}. In this
section, we briefly describe the technologies that enable the disruptive
performance improvements of 5G.

\textbf{Millimeter Wave Spectrum.} In addition to the classical spectrum below 6
GHz used by the majority of wireless communication technologies, 5G will operate
in a high-frequency spectrum, from 28 GHz up to 95 GHz \cite{5g_ch_opp,
5g_5tech}. This range is known as the \textit{millimeter wave (mmWave)
spectrum}. Compared to previous cellular network technologies, 5G will use a
larger band of frequencies, thus, avoiding congestion. In comparison, 4G
operates typically in the range 700-2600 MHz \cite{5g_ch_opp,wiki_lte_bands}.

\textbf{Massive MIMO and Beamforming.} 5G uses the \textit{massive
multiple-input and multiple-output (MIMO)} technology \cite{mimo}. This
technology consists of large antenna formations in both the base station and the
device to create multiple paths for data transmission. With massive MIMO
technology, 5G can achieve high spectral efficiency \cite{5g_ch_opp} and better
energy efficiency \cite{mimo}. Beamforming\footnote{The terms beamforming and
massive MIMO are sometimes used interchangeably \cite{beamforming}.} is a subset
of massive MIMO \cite{5g_nr_mimo}. Beamforming controls the direction of a
wave-front by manipulating the phase and magnitude of the signals sent by a
single antenna placed in a formation of multiple antennas. In this way,
beamforming identifies the most efficient path to deliver the data to a
receiver, while reducing the interference with nearby terminals. In addition, 5G
uses a full-duplex technology which doubles the capacity of wireless links at
the physical layer. With full-duplex, a device is able to transmit and receive
data at the same time, using the same frequency \cite{full_duplex}. Based on
these new technologies, it is predicted that 5G has the potential to improve
services at the edge, support more use cases, accelerate the development of
smart cities, and enhance user experience \cite{5g_nr_mimo}.

\textbf{Small Cells.} In addition to a larger spectrum and massive MIMO, 5G will
comprise densely distributed networks of base stations in \textit{small cell}
infrastructure. This enables enhanced mobile broadband (eMBB) and low
latency~\cite{nokia_smallcell}, providing an ideal infrastructure for edge
computing. While small cells are typically used to cover hot spots, in mmWave
5G, they become a necessity due to the high-frequency (above 28 GHz) radio waves
that cannot cover the same area as the classical low frequencies (below 6
GHz)~\cite{nokia_smallcell}.

\begin{table}[tp]
\centering
\caption{A comparison of the specifications of mobile communication technologies}
\label{tab:mobile_comm_comp}
\resizebox{0.489\textwidth}{!}{
\begin{tabular}{|l|r|r|r|r|r|}
\hline
& \multicolumn{1}{c|}{\multirow{2}{*}{\textbf{5G}}} &
\multicolumn{1}{c|}{\multirow{2}{*}{\textbf{4G}}} &
\multicolumn{1}{c|}{\textbf{WiFi}} & \multicolumn{1}{c|}{\multirow{2}{*}{\textbf{Bluetooth 5}}} \\
& & & \multicolumn{1}{c|}{(802.11ac)} & \\
\hline
\hline
Bandwidth [Gbps] & 10-20 & 1 & 0.4-7 & 2 Mbps\\
Latency [ms] & 1 & 10-100 & 0.9/6.2 & 200 \\
% Range [m] & & & 50-100 & 10-240 \\
Mobility [km/h] & 500 & 350 & - & - \\
Frequency [GHz] & 0.6-6, 28-95 & 0.7-2.6 & 5 & 2.4 \\
% \multirow{2}{*}{Connected Devices} & 1,000,000 & 100,000 & 200 & 7
% \\ & per km\textsuperscript{2} & per km\textsuperscript{2} & per gateway &
% per gateway \\
Connected Devices & 1,000,000 / km\textsuperscript{2} & 100,000 / km\textsuperscript{2} & 200 / gateway & 7 / gateway \\
Year & 2019 & 2009 & 2014 & 2016 \\
\hline
\end{tabular}
}
\end{table}

\begin{table*}[tp]
\centering
\caption{Measurements on current 5G deployments}
\label{tab:5g_deployment_speed}
\resizebox{0.87\textwidth}{!}{
\begin{tabular}{|c|c|c|c|c|c|c|}
\hline
\multirow{2}{*}{\textbf{Location}} & \multirow{2}{*}{\textbf{Date}} &
\multirow{2}{*}{\textbf{Operator}} & \multirow{2}{*}{\textbf{Device}} &
\textbf{Bandwidth} & \textbf{Latency} & \multirow{2}{*}{\textbf{Reference}} \\
& & & & (max) [Mbps] & (min RTT) [ms] & \\
\hline
\hline
Chicago, USA & 19/5/2019 & Verizon & Samsung Galaxy S10 5G & 1,385 & 17 & \cite{5gtest_verizon_chicago} \\
Chicago, USA & 30/6/2019 & Verizon & Samsung Galaxy S10 5G & 1,070 & - & \cite{5gtest_tomsguide} \\
New York, USA & 30/6/2019 & T-Mobile & Samsung Galaxy S10 5G & 579  & - & \cite{5gtest_tomsguide} \\
New York, USA & 1/7/2019 & T-Mobile & Samsung Galaxy S10 5G & 529 & 53.5 & \cite{5gtest_tmobile_newyork} \\
\hline
Bucharest, Romania & 19/7/2019 & RCS/RDS & Xiaomi Mi Mix3 5G & 458/20.6 & 12
& this paper \\
Bucharest, Romania & 6/9/2019 & RCS/RDS & Huawei Mate 20 X 5G & 458/8.5 & 6
& this paper \\
\hline
\end{tabular}
}
\end{table*}

\textbf{Device-to-device Communication.} Similar to Bluetooth and WiFi (i.e.
WiFi-Direct), 5G allows devices to communicate with each other directly, with
minimal help from the infrastructure \cite{Biswash2017}. This
\textit{device-to-device} (D2D) communication is a key feature of 5G that has
the potential to accelerate the development of edge-centric applications. For
example, in automotive applications, vehicles will be able to talk directly to
each other, thus, reducing latency and avoiding the failure of the connection to
the base station. Other use cases of D2D 5G communication are federated
learning, where edge devices could share data among them, and blockchain where
devices need to establish peer-to-peer (P2P) connections.

\textbf{Virtualization.} In addition to the improvements in the physical layer,
5G networks are going to be highly virtualized. Among the virtualization
technologies used by 5G, we distinguish software-defined networking (SDN),
network function virtualization (NFV), and network slicing. SDN is an approach
that separates networking data plane (i.e. data forwarding process) from the
control plane (i.e. the routing process). This separation leads to easier
configuration and management, and higher flexibility and elasticity
\cite{sdn_survey}. Complementary to SDN, NFV \cite{nfv_survey} uses commodity
hardware systems to run networking services that are traditionally implemented
in hardware, such as routers and firewalls. With NFV, network flexibility is
greatly improved, and the time-to-market is reduced, at the cost of lower
efficiency compared to dedicated hardware.

\textbf{Network Slicing.} Based on SDN and NFV, 5G networks will employ
\textit{network slicing} to multiplex virtualized end-to-end networks on top of
a single physical infrastructure. By separating infrastructure operators and
service providers, 5G will better utilize hardware resources while providing a
diversified range of services to both businesses and end-users
\cite{nslicing_survey}. However, all these virtualization technologies pose new
challenges in terms of security management and monetization, as we shall analyze
in this paper.

\textbf{Perfomance Improvements.} Compared to 3G and 4G, 5G has a lower latency
of approximately 1 ms, increased energy efficiency, and a peak throughput of
10-20 Gbps \cite{5g_ch_opp,min_int2020}. The increase in bandwidth will not only
support better user experience, but also allow for more connected devices, such
as drones, vehicles, and AR goggles, among others. While a 4G base station can
only support around 100,000 devices, 5G can support up to a million devices per
square kilometer~\cite{min_int2020}. A 5G network is designed to be flexible and
suited for edge deployment, which further improves the end-to-end latency and
overall user experience.

In the context of IoT devices and their use cases, we compare the specifications
of major wireless communication technologies in \autoref{tab:mobile_comm_comp}.
5G has almost always the best characteristics, except for latency, where newer
generations of WiFi have similar specifications.
However, median WiFi latency on an 802.11n router is 0.9 ms and 6.22 ms, when 5
GHz and 2.4 GHz frequencies are used for measurements \cite{hpbn_book_14},
respectively. The 99th percentile goes up to 7.9 ms and 58.9 ms for the two
frequencies \cite{hpbn_book_14}, respectively. In practice, current 5G
deployments exhibit latencies in the range of 6-40 ms, with jitter than goes up
to 145 ms, as shown in Appendix~\ref{sec:apdx_5g}.

\subsection{Current 5G Deployments}
\label{sec:5g_deployments}
% Dumi

In August 2019, some countries and operators were offering commercial 5G
networks, with limited deployment. According to a map published by
Ookla\footnote{Ookla developed \url{speedtest.net}, a widely-used tool to
measure the speed of Internet connections in terms of download and upload
throughput, and ping latency. In this paper, we use Ookla's Speedtest Android
application for measurements.} \cite{5g_map}, only the USA and Uruguay have 5G
networks in the Americas. Besides Uruguay, only South Africa and Australia have
5G in the southern hemisphere. In Asia, 5G is available in some countries in the
Middle East, such as Saudi Arabia, Qatar, Kuwait, and the United Arab Emirates,
as well as in South Korea. In Europe, 5G is available in the UK, Spain, Germany,
Switzerland, Italy, Romania, and Finland.

\autoref{tab:5g_deployment_speed} summarizes existing speed tests on 5G networks
around the world. All these tests are conducted using Ookla's Speedtest Android
application. At the bottom of this table, we present our tests done in
Bucharest, Romania where there is a 5G network provided by the local operator
RCS/RDS using base stations produced by Ericsson \cite{5g_digi_ericsson}.
These tests were done with two smartphones equipped with 5G modems, (i) a Xiaomi
Mi Mix3 5G (\textbf{Mix3}) \cite{mix3} and (ii) a Huawei Mate 20 X 5G
(\textbf{MateX}) \cite{matex}. The former device has a Qualcomm X50 5G modem,
while the latter is equipped with Huawei's own 5G modem.

We analyze the best results in this section, and present the details in
Appendix~\ref{sec:apdx_5g}. Both smartphones exhibit a maximum download
throughput of 458 Mbps on 5G, compared to a maximum of 86.3 Mbps on 4G. But the
upload throughput of 5G is similar or even lower compared to 4G. For example,
the maximum upload throughput on 5G is 20.6 and 8.5 Mbps with Mix3 and MateX,
respectively. On 4G, the maximum measured upload throughput is 23.9 and 29.6
Mbps with Mix3 and MateX, respectively. The latency is measured in terms of
round-trip time (RTT), similar to the results reported by the \textit{ping}
Linux tool. RTT is accompanied by jitter, representing the deviation from the
average latency. The minimum RTT measured on 5G with Mix3 and MateX is 12 and 6
ms, respectively. These RTTs have corresponding jitters of 4 and 1 ms on Mix3
and MateX, respectively. The RTT on 4G is much higher, with a maximum of 48 and
54 ms on Mix3 and MateX, respectively. On average, the download throughput of 5G
across different test servers and measured in different locations is above 300
Mbps, while the upload speed is always less than 30 Mbps.

While the download speed\footnote{We use the terms bandwidth and speed
interchangeably.} is significantly higher compared to 4G, both the upload speed
and latency are surprisingly low, compared to the
specifications~\cite{min_int2020}. This is caused by two main factors. First,
the speed test consists of sending requests to servers that are not very close
to the base station. Hence, the latency and throughput are influenced by both
(i) the 5G wireless link to the base station and (ii) the wired, optic, wireless
or mixed path from the base station to the server. Second, current 5G setups are
using the Not-Stand-Alone (NSA) mode \cite{5G_nsa_ericsson}. Only the
Stand-Alone (SA) 5G mode is supposed to achieve an ultra-low latency of 1 ms
\cite{5G_nsa_ericsson}. In terms of upload throughput, network operators tend to
limit it because typical users are affected more by the download speed.

We note that our measurements are influenced by the connection between the base
station and the test server. In our case, the number of hops between the device
and the test server is in the range 9-13. While the number of hops is relatively
high, the backhaul links connecting the base stations to the Internet are of
high capacity. These links are usually based on fiber or microwave technology
with capacities of up to 20 Gbps \cite{5gbackhaul}. On the other hand, these
measurements expose pertinent download and upload throughput as experienced by
the end-user.

\subsection{5G Simulators}
\label{sec:simulation}

% Why simulators
With the limited amount of both 5G deployments and 5G-ready terminals, it is
mandatory to explore simulation and emulation solutions for 5G in an effort to
develop and analyze applications targeting this new technology. Since our focus
is on the impact of 5G on data-driven software platforms, the simulator should
be able to reproduce networking behavior at a high level, in terms of
throughput, latency, jitter, and packet loss.

% Existing approaches
Many 5G simulators, such as MATLAB 5G Toolbox~\cite{5g_matlab},
NYUSIM~\cite{NYUSIM_17}, NetTest 5G Network
Emulators\footnote{\url{http://www.polarisnetworks.net/5g-network-emulators.html}},
ns-3\footnote{\url{https://www.nsnam.org/}}, \cite{e2e_simulator}, focus on the
physical layer (i.e. radio access network - RAN). Such a detailed simulation at
the physical level could offer useful insights to network engineers and mobile
operators, but it is time-consuming and resource-intensive.

% Our approach
Instead, simpler solutions could be used to emulate a 5G environment. For
example, the \textit{tc} Linux
tool\footnote{\url{https://linux.die.net/man/8/tc}} is able to introduce delays
with different patterns to emulate higher latency with custom jitter
distributions. Moreover, \textit{tc} can limit the bandwidth of a given
interface. One can use \textit{tc} in a cluster with Gigabit Ethernet or higher
bandwidth links to emulate 5G networking conditions. However, it remains to be
investigated how to emulate both D2D and device-to-base station communications
on top of an Ethernet network.

% Mininet
In conjunction with network virtualization, which is a key feature of 5G,
existing solutions for quick prototyping with SDN can be used, such as Mininet
\cite{mininet_10} and its fork, Mininet-WiFi \cite{fontes2015mininet}.
Some researchers explored the idea of using Mininet as a platform to emulate 4G
and 5G on top of wired or wireless networks
\cite{5g_emulation_springer,5g_mininet_sdn,4g5g_emulation}. However, some of the
results reported by these projects are far from both the specifications of 5G
and our preliminary measurements. For example, the throughput reported in
\cite{5g_mininet_sdn} is below 100 Mbps. Hence, special care needs to be adopted
when using these prototyping platforms to conduct performance measurements for
data management and processing frameworks.

%%% Anh

%%%%ooibc: Anh, can this be classified as serverless processing?

\section{Areas of Impact}
\label{sec:impact}
%% all

\begin{figure}[tp]
\centering
\includegraphics[width=0.48\textwidth]{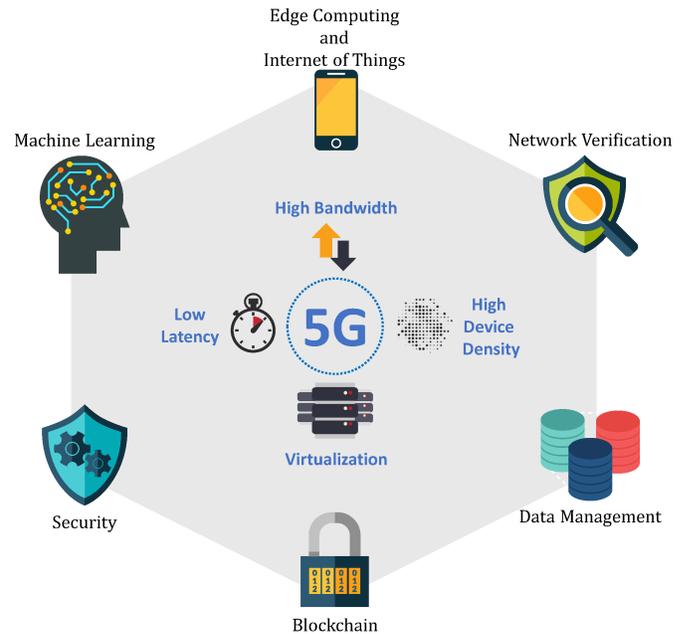}
\caption{Areas impacted by 5G}
\label{fig:impact_full}
\end{figure}

5G and its revolutionary features are going to impact multiple computer science
domains and create new use cases, as highlighted in \autoref{fig:impact_full}.
In this section, we present our view on the domains impacted by 5G, such as edge
computing, database systems, artificial intelligence, security, among others. In
the next section, we present some use cases where 5G and related technologies
will have a significant impact.

\textbf{Motivating Use Case.} Before diving deeper into each area impacted by
5G, we motivate our analysis with a use case that covers all the areas mentioned
and shows the relationship among them. Representing the growing market of
healthcare (\cref{sec:healthcare}), this use case assumes that patients
(end-users) take ownership of their medical data, also known as electronic
health records (EHR) or electronic medical records (EMR). This represents a
global trend that tries to put the patient at the center of the healthcare
system. For example, Apple allows users to download and keep their medical
records on the iPhone \cite{apple_ehr}.

\begin{figure}[tp]
\centering
\includegraphics[width=0.485\textwidth]{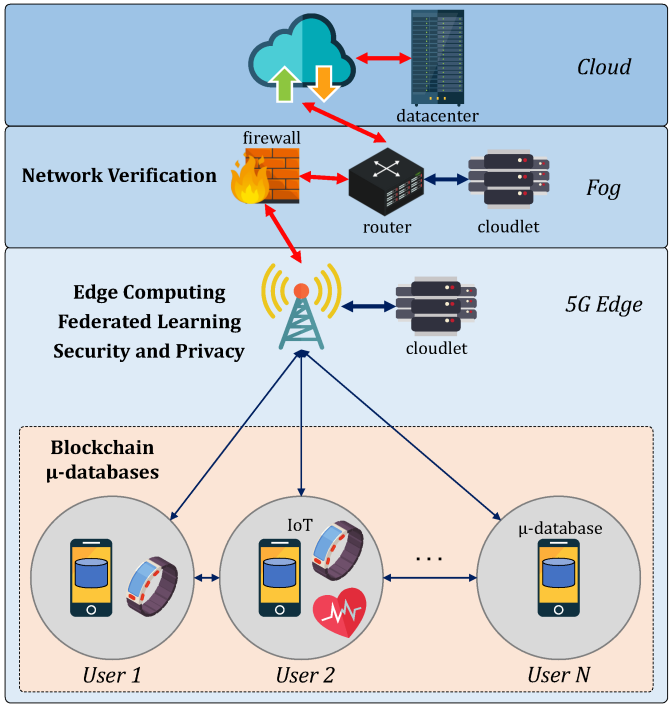}
\caption{Motivating use case}
\label{fig:usecase}
\end{figure}

By storing medical records locally, the user's smartphone becomes a small
database that we refer to as \textit{$\mu$-database} (\cref{sec:db}), as shown
in \autoref{fig:usecase}.
In addition to medical records downloaded from clinics and hospitals, this
$\mu$-database stores data collected by a variety of IoT devices, such as a
smart watch and mobile electrocardiogram. Within a 5G network, these
$\mu$-databases could be interconnected either through the base station or
directly, using the D2D feature of 5G. Furthermore, the emerging blockchain
technology can be used to improve the security of these $\mu$-databases in a
hostile environment.

These data could be used to train medical deep learning models, such as disease
progression models \cite{Kaiping_DPM_17}, which are then distributed to the
devices to analyze new data and to send alerts to doctors. To make the training
more efficient, federated learning (\cref{sec:ai}) is used to distribute the
work among multiple devices with the help of a coordinator. With the emergence
of edge computing (\cref{sec:edge}), the coordinator could be placed at the
edge, in a virtualized micro-datacenter or cloudlet \cite{cloudlets_2009}.

Nevertheless, the biggest concern in modern healthcare is the security and
privacy of the patients' data. Recent data breaches \cite{uk_data_breach,
sg_data_breach} motivate the research of new security techniques.
With the adoption of 5G, its virtualization feature could help in isolating the
healthcare use case from other verticals. However, virtualization does not
always ensure privacy (\cref{sec:security}). In addition, security management is
problematic in the presence of network slicing, and both the physical
infrastructure and SDN configurations need to be verified to ensure a secure
environment (\cref{sec:netverif}).

Another method to increase the security and privacy of distributed
$\mu$-databases is the use of blockchain technology (\cref{sec:blockchain}). For
example, some startups, such as MediLOT \cite{medilot} and
Medicalchain \cite{medicalchain}, propose patient-centric healthcare based on
blockchain. However, the scalability of blockchain \cite{anh_blockchain_tkde}
remains an open problem in the context of 5G networks.

% %% Dumi
\subsection{Edge Computing and the Internet of Things}
\label{sec:edge}

\subsubsection{Overview}

Edge computing is a relatively new paradigm that proposes to move cloud services
closer to the users and to the devices that produce data, at the edge of the
network \cite{edge_vision_2016}. With a growing number of devices connected to
the Internet \cite{IoTAnalyticsDevices2018}, the pressure on the backbone links
of the Internet is increasing. Edge computing alleviates this issue by
performing some or all computations closer to the devices that produce data.
Depending on the location of these computations, we distinguish between edge and
fog computing. In edge computing, the processing is done on the device or one
hop away from the device, for example in a mini-datacenter connected to the 5G
base station \cite{edge_mec_survey_2017}, as shown in \autoref{fig:usecase}. On
the other hand, in fog computing, introduced by Cisco in 2015
\cite{cisco_fog_2015}, the computation could be done anywhere between the edge
and the cloud, in switches, routers, base stations or other networking devices.

Edge computing has multiple flavors, among which we distinguish (i) the fog,
(ii) Multi-access Edge Computing (MEC), and (iii) cloudlets
\cite{edge_vision_2016,edge_mec_survey_2017}. The fog is an extension of the
edge, where the processing can be done on the way to the cloud, in the
backbone's switches and routers. MEC is a set of standards addressing the
diversity of protocols, applications, services, and providers of edge computing.
At its core, MEC is based on virtualization to provide cloud-like services at
the edge, within the range of RAN~\cite{mec_etsi}. A cloudlet
\cite{cloudlets_2009} is a small datacenter connected to a networking access
point. Typically connected to a base station, one hop away from the devices, a
cloudlet can also be placed in the fog, as shown in \autoref{fig:usecase}. A
cloudlet is using virtualization to provide computing and storage services.
Thus, cloudlet technology can be viewed as part of MEC. With 5G being a heavily
virtualized technology, we are expecting an accelerated deployment of MEC and
cloudlets.

With the adoption of 5G, which enables higher bandwidth and more connected
devices compared to 4G, edge computing becomes a necessity because current cloud
interconnections are not able to sustain the traffic. Our measurements on Google
Cloud Platform (GCP) and Amazon Elastic Compute Cloud (EC2) show that bandwidths
between different cloud datacenters (regions) can hardly hit 100 Mbps, while the
majority of our measurements are below 10 Mbps. Only closely located regions,
such as those in Western Europe, exhibit bandwidths of up to 92.8 and 126 Mbps,
for GCP and EC2, respectively. These bandwidths are far from being able to
sustain the demands of 5G edge devices, where a single device could upload
data with a throughput of up to 1 Gbps \cite{min_int2020}.

The pressure on the backbone and cloud inter-region links is going to increase
as 5G is considered to be the network for IoT \cite{5g_ch_opp}. In current
deployments, IoT devices typically connect to a gateway or the Internet through
WiFi, Bluetooth, Long Range (LoRA), Zigbee, among others \cite{MEKKI20191}.
These protocols are suitable for short-range communications with low mobility,
used in applications such as smart homes and smart offices.
However, they may not be suitable for larger deployments, such as smart cities
and smart farms, and high-mobility applications such as automotive. Next, we
identify a series of devices that can benefit from 5G features, as listed below.

\begin{itemize}

\item \textbf{Surveillance systems}. These systems allow users to remotely scan
the area inside or around their homes from the comfort of a smartphone.
They can see who is snooping around while overseas, and alert the authorities if
needed. For such a service, there is a need for good video quality and high
frame rate. 5G has enough bandwidth to deliver high-quality UHD video with low
delay while allowing flexible reconfiguration which is not possible when using
wired connections. On the other hand, edge computing helps in pre-processing the
image stream in a cloudlet and sending only the alerts to the cloud.

\item \textbf{Autonomous cars}. Undoubtedly, smart cars need fast response
times. The theoretical latency improvement from 50 ms in 4G to 1 ms in 5G may be
enough to avoid accidents. Moreover, the D2D communications in 5G will have a
positive impact on Vehicle-to-Vehicle (V2V) messaging, further reducing the
latency compared to going through a base station. Using D2D, automotive
communications can avoid the problem of a single point of failure represented by
a faulty base station. As such, we predict that smart cars will adopt 5G to
deliver large volumes of data with high speed to avoid accidents.

\item \textbf{Drones}. In emergencies and dangerous situations, such as
search-and-rescue, firefighting, surveying, delivery services, having high
network bandwidth allows the drones to send high-quality sound and video back to
the command center, at the edge. The low latency of 5G allows better control
over the drone compared to 4g or WiFi. Similar to autonomous cars, smart drones
could benefit from D2D communications, especially in remote areas with no access
to a base station. Using D2D communications, a group of closely located smart
drones can form a swarm to work towards a common objective \cite{shrit2017new}.

\item \textbf{Healthcare devices}. With a high bandwidth and low latency, 5G
would improve the monitoring of patients with chronic diseases. Vital signs can
be sent to the doctor or hospital with high frequency, while alerts can be
triggered as soon as the edge device detects something wrong.

\end{itemize}

Some of the features of 5G address the challenges faced by the IoT domain.
First, the increasing number of IoT devices could be handled by the superior
device densities supported by 5G. According to a survey by IoT Analytics
Research, there were 7 billion IoT devices in 2018
\cite{IoTAnalyticsDevices2018}. This number will triple by 2025
\cite{IoTAnalyticsDevices2018}. This increase in the number of IoT devices
raises concerns about connectivity and security, among others. According to an
online survey of IoT development conducted by the Eclipse Foundation in 2019
with 1717 participants \cite{EclipseIoTSurvey19}, the top three concerns are
security (38\%), connectivity (21\%), and data collection and analytics (19\%).

\subsubsection{Challenges and Opportunities in Edge Computing}

% 1: diversity at the edge -> hard to implement -> virtualization could solve it
Nonetheless, some challenges need to be considered and analyzed carefully before
the successful adoption of 5G in edge computing. First, there is a high
diversity of edge hardware, communication protocols, service providers and
processing frameworks. To overcome this, the European Telecommunications
Standards Institute created a special group of interest to propose standards for
MEC \cite{edge_mec_survey_2017}. 5G could address this issue using
virtualization, where hardware functions and software protocols can be
virtualized on commodity infrastructure, thus, decreasing the prototyping and
deployment time.

% virtualization
Edge services could be driven by the 5G end-to-end network virtualization. For
example, network slicing would allow different application planes to run in
isolation on the same infrastructure, as shown in \autoref{fig:net_slices}.
With network slicing, cloud-like services at the edge, the broadband
connectivity plane, and smart city applications could all run in isolation.
Nevertheless, the security of such a setup is challenging, as we shall see in
the next sections.

% 2: remote management
A second factor that hinders 5G edge adoption is the high cost of installing,
protecting and maintaining edge devices in remote areas \cite{roman2018mobile}.
5G is able to address these issues partially by employing its high bandwidth and
low latency features. The former implies that advanced, high-definition security
monitoring solutions can be deployed together with the edge hardware. The latter
helps in detecting and acting on problems faster, from a centralized command
facility.

% 3: energy
A third factor that needs to be considered in 5G edge computing is energy
efficiency. Remote edge devices may face energy constraints due to the lack of
connections to the power grid. Operating on alternative sources of energy, such
as solar panels or batteries, imposes constraints on both computation and
communication. While communication is often more energy-expensive compared to
computation, it is a challenge to decide when to process the data at the edge
and when to offload it to a cloudlet \cite{edge_vision_2016}. With its superior
energy efficiency, 5G could help in improving the overall efficiency of edge
computing.

\subsubsection{Challenges and Opportunities in IoT}

% Security
One of the biggest challenges IoT development has to encounter in a 5G
environment is the new wave of security threats. Since the number of connected
IoT devices continues to rise, there is a higher risk that systems will be
attacked by malware and ransomware to steal sensitive data or to perform DDoS
attacks \cite{ddos_botnet}. This problem is more stringent with the IoT devices
being used in automation and security systems at home or in vehicles. These
systems may be compromised, leading to more serious threats, such as home
intrusion or remote vehicle hijack. A piece of common advice to ensure the
security of IoT devices is to keep their firmware and security patch up-to-date
to avoid any vulnerabilities exploitable by attackers. Also, users need to
change their default account and password periodically on the IoT devices to
prevent unauthorized access by brute-force attacks. Finally, data transmission
and communication between devices or from the devices to the 5G network need to
be encrypted to prevent any leak of confidential data.

% Data privacy
The second challenge that needs to be considered is how to guarantee data
privacy when IoT devices have access to private data such as surveillance
videos, daily habits, and health data. Although users can review the
coarse-grained access control of these devices to sensitive information
\cite{roman2018mobile}, they generally cannot supervise how these data are
collected and utilized. For example, if users let IoT applications access their
data, they will not know when the devices send the data to developers,
advertisers, or any other third parties. To ensure the privacy of data, multiple
solutions need to be considered. For example, it is important to set dedicated
policies, regulations, rules, or laws to ensure that IoT service providers and
developers take necessary actions to protect users' sensitive data. Besides,
strong security solutions also need to be set up in IoT devices, to prevent any
breach or exploitation of sensitive data.

% Bugs
The third challenge is that most IoT devices might not be tested sufficiently
during their production and might not receive enough firmware updates after
deployment. This might be due to the fierce competition in the IoT industry
where manufacturers often focus on quickly producing and selling devices and do
not pay enough attention to security issues. For example, many manufacturers
only offer firmware updates for new devices, while stopping the update of
old-generation devices when they start working on the new generation. This bad
practice might leave IoT devices vulnerable to potential attacks due to the
outdated firmware. To overcome this problem, manufacturers are encouraged to
test their products properly and update firmware regularly. Furthermore, they
should use safe programming languages and automatic program testing and
verification techniques to avoid potential bugs during product development.

% Monetization
A fourth factor is the cost of a 5G subscription per IoT device. Given the large
count of IoT devices, a linear pricing scheme is not going to incentivize the
adoption of 5G in IoT. To leverage this, hybrid deployment could be used, where
multiple IoT devices connect to a 5G gateway using traditional protocols, such
as WiFi or LoRA.

\subsection{Network Testing and Verification}
\label{sec:netverif}

\begin{figure}[tp]
\centering
\includegraphics[width=0.485\textwidth]{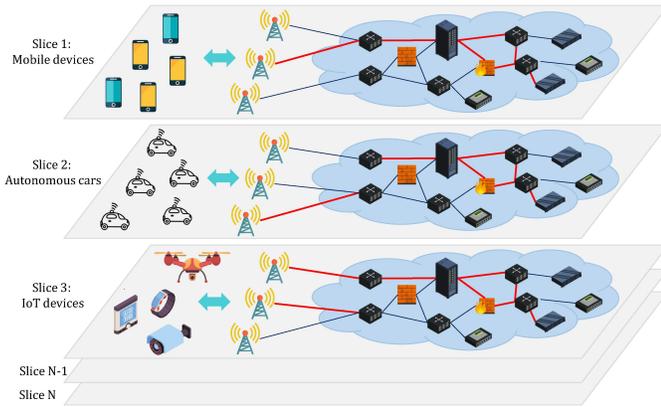}
\caption{5G network slices}
\label{fig:net_slices}
\end{figure}

\subsubsection{Overview}

Ensuring that modern networks, including 5G, operate properly as designed is
crucially important to telcos, banks, content providers and other businesses.
The failure of these networks might lead to severe consequences. For example,
according to a report of IHS Inc. in 2016 \cite{NetworkDowntime}, network
failure caused the loss of billions of US dollars annually in North American
businesses. The failure of a network can occur statically due to its
misconfiguration or dynamically at runtime \cite{ZengKVNNetworkTesting}. The
misconfiguration errors are often introduced by human mistakes, and they can
lead to problems such as unreachable servers, or security holes. On the other
hand, runtime errors are due to failures in network links and hardware, or bugs
in network software.

Huge efforts have been spent by researchers to develop testing and verification
techniques to find network errors \cite{LiYWYSWZW19}. However, this task is
known to be very complex since modern networks include a large number
(thousands) of not only servers, routers, and end-user devices, but also many
middleboxes such as firewalls, load balancers, transcoders, proxies, and
intrusion detection systems \cite{PandaASSS15}. Furthermore, the software
controlling these devices is very complicated: it contains millions of lines of
code and runs in a highly distributed environment. For 5G, this testing and
verification task will be more challenging, due to the network's growth in
complexity and flexibility. In particular, a 5G network can support a massive
number (up to millions) of connected devices. It is also equipped with a novel
network slicing feature which allows a slice (or virtual network) to be
dynamically created, used, and deleted. In the following, we will discuss more
about such difficulties and challenges.

\subsubsection{Network Testing}

Up to date, testing is the main method that has been used to discover
errors in modern networks \cite{NetworkTestingSurvery}. Network engineers
often find bugs by using a wide range of tools, from the rudimentary
\texttt{ping} and \texttt{traceroute}, to advanced tools like
\texttt{nmap}, \texttt{tcpdump}, \texttt{netcat}, \texttt{acunetix},
\texttt{ip scanner}. They mostly conduct ad-hoc validations of existing
networks (3G, 4G, or enterprise networks) via active monitoring to detect
potential problems. For example, a network often needs to be validated
after a configuration change, such as when new remote sites are installed,
routing policies are changed, or firewall rules are updated.

However, the existing network testing solutions might need to evolve to cope
well with the scale of 5G. Unlike 3G and 4G networks, which are limited only to
the telecom industry, a 5G network will comprise millions of connected devices
from many industry verticals \cite{nslicing_survey} grouped in network slices,
as shown in \autoref{fig:net_slices}. Therefore, various ad-hoc cases of the
network need to be considered for validation. Also, monitoring the network will
be more challenging since network slices can be flexibly created, used, and
deleted, based on user requirements \cite{nslicing_survey}. Hence, it is
difficult for network engineers to manually design testing strategies that
thoroughly cover all behaviors of the network.

In order to make 5G network testing more effective and efficient, the tasks need
to be automated as much as possible. Firstly, there is a need to design tools
that can automatically analyze the network configuration and generate tests to
cover all ad hoc cases. Secondly, these test cases need to be run automatically
and periodically on candidate networks to discover any possible errors.
Although automatic testing is new in the context of network testing, this idea
has been well-studied by the software engineering community.
Recently, several efforts have been made to automate network testing.
For example, Zeng et al.~\cite{ZengKVM12} have proposed a technique to
automatically generate test packages for testing the forwarding behavior of
simple networks. However, the size and dynamicity of 5G will be much more
complicated than existing networks. Hence, there are many challenges and
opportunities to develop automatic testing tools for 5G networks.

\subsubsection{Network Verification}

Although testing is commonly used to detect network errors, this method has two
limitations. Firstly, it is often used to check the behaviors of a production
network but not to examine the network's configuration. Secondly, this approach
cannot guarantee that a network is implemented correctly according to its design
since it is impossible to generate test cases that cover all the possible
behaviors of the network. In reality, it is often desired that a network's
configuration can be examined before being deployed to prevent any possible
errors in the future.

Inspired by recent advances in software verification, researchers have proposed
to treat networks like programs so that they can apply similar techniques to
formally verify the forwarding behavior of networks \cite{PandaASSS15,
Varghese15}. In essence, a network consists of two planes, namely, a \emph{data
plane} and a \emph{control plane} \cite{kurose_networking_book}. The data plane
decides how a network packet is handled locally by a router: when the packet
arrives at one of the router's input links, it will determine which output link
to forward it to. The control plane determines how a packet is routed among
routers along a path from the source node to the destination node.
In traditional networks, both these planes are implemented in routers.
However, in modern networks, the control plane can be implemented as a separate
service in centralized servers.

Recent works have focused on verifying simple networks, which are configured by
static and immutable forwarding rules \cite{KazemianVM12}, or small-scale
networks with a limited number (hundreds or thousands) of devices
\cite{BeckettGMW17, DobrescuA14, KhurshidZZCG13}. In reality, modern networks
often contain various middleboxes, whose states are mutable and can be updated
in response to received packets. Hence, the behaviors of routers and middleboxes
in these networks are affected by not only their configurations but also by the
incoming packets. Further, 5G networks will be even more complex since they
allow a massive number (millions) of connected devices. These two challenges,
namely, complexity and mutable states, will be the key factors that need to be
considered when verifying the forwarding behavior of 5G networks.

As previously mentioned, 5G supports network slicing, where a network slice is a
software-based, logical network that can span across multiple layers of the
network and could be deployed across multiple operators.
Furthermore, the isolation of slices can be flexibly configured at different
levels to satisfy the customers' needs \cite{nslicing_survey}.
For example, some users may not mind sharing network resources with others but
would require isolation for the computing resources. Therefore, it will be
challenging to formally verify if the isolation property of network slices in a
deployed network satisfies its design.

%%% Dumi & Ooibc & Lin Qian & Xiaokui
\subsection{Data Management and Processing}
\label{sec:db}

\subsubsection{Overview}

5G could be the P2P network layer in a system comprising millions of
interconnected $\mu$-databases. A $\mu$-$database$ stores a subset of the data
corresponding to a certain application. For example, a medical $\mu$-database
stores a part of the patient's medical records, where the entire dataset is
represented by all the records of all the patients using the same medical
application or going to the same group of hospitals. With a projected density of
one million devices per square kilometer~\cite{min_int2020}, 5G is able to
connect a few million devices in a smart city, where each device stores a
$\mu$-database. This either gives rise to (i) a network of interconnected
$\mu$-databases or (ii) millions of independent $\mu$-databases owned by
individuals.

The realization of the first scenario in a traditional datacenter equipped with
high-performance server systems is problematic due to networking and power
constraints. Firstly, the connectivity in a datacenter's cluster is done through
switches or routers that become either bottlenecks or sources of network
failure. Secondly, a typical server uses more than 50 W of power, while often
reaching 100 W \cite{Loghin_VLDB_15}. With one million servers, the power
requirement of such a datacenter reaches 100 MW, $10\times$ more than the
fastest supercomputer in Top500 \cite{top500_jun19}.

On the other hand, low-power systems based on ARM CPU, such as smartphones and
IoT devices, typically use less than 10 W when active \cite{Loghin_EDGE_17,
Loghin_VLDB_15}. Besides, 5G is predicted to be more energy-efficient
\cite{5g_ch_opp}, hence, it will further reduce the power usage of the node.
Previous research projects connecting low-power nodes in distributed data
management and processing systems \cite{tarazu_asplos12, fawn_sosp09,
Loghin_PEVA_2015, Loghin_VLDB_15} show that these devices can significantly
reduce energy usage, while trading-off performance in terms of response time and
throughput. Indeed, there is a high possibility that a distributed network of
more than one million low-power 5G devices will exist in the near future.

The second scenario is that of a P2P database comprising millions of
$\mu$-databases, where each individual user owns their data, stores them on its
own device, and decides how to share them. For example, a user can store her
entire medical history on the smartphone, instead of keeping fragmented records
in the databases of different hospitals. By storing the data locally, the user
has better control on privacy and sharing. In addition to data, the user may
choose to share resources, such as storage space or computing units. In this
scenario of independent $\mu$-databases, sharing requires a fine-grain access
control mechanism to ensure security and privacy, especially in the context of
strict data protection and privacy rules, such as the European Union's General
Data Protection Regulation (GDPR) \cite{gdpr}.

\subsubsection{Challenges and Opportunities}

% %% GDPR
With the ownership of the data being passed back to the users and with the
implementation of strict data protection frameworks, such as GDPR, big data
analytics has to be efficiently supported both on the cloud and at the edge.
First, it is challenging to perform efficient batch or stream processing in the
presence of a fine-grain access control mechanism. For example, users may choose
to share only part of the data which could affect the final results of the
analytics task. Second, the highly dispersed and volatile nature of distributed
P2P $\mu$-databases make fault-tolerance and task scheduling stringent issues in
big data processing frameworks. It is well-known that strugglers affect the
performance of data analytics \cite{Zaharia_Hadoop_08, bigdata2015}.

%%% P2P and data movement and resource sharing
When large amounts of fragmented data are stored across a large number of
devices, data have to be processed locally and/or transferred to the cloud for
large scale analytics. For local data management and resource sharing over
devices, efficient and light data management is required, possibly with some
form of distributed shared memory \cite{Cai_RDMA_18}.

% %% human-in-loop -- AI, crowdsourcing etc
The D2D and high density of connections in 5G will present a great opportunity
for human-in-loop data processing. A complex high-level job may be partitioned
into computer-based tasks and human-based tasks \cite{cdas}, which is in line
with the exploitation of AI for tasks that machines can do best.
Decomposition and classification of tasks need to be designed for specific
application domains since domain knowledge and availability of experts are key
to high accuracy. Fast response from a human is needed to improve the overall
data quality and decision-making. However, all these must be examined to ensure
that humans do not introduce noise into the system and cause further
irregularity.

%%% Dumi
\subsection{Blockchain}
\label{sec:blockchain}

\begin{figure}[tp]
\centering
\begin{subfigure}{0.485\textwidth}
\centering
\includegraphics[width=0.75\textwidth]{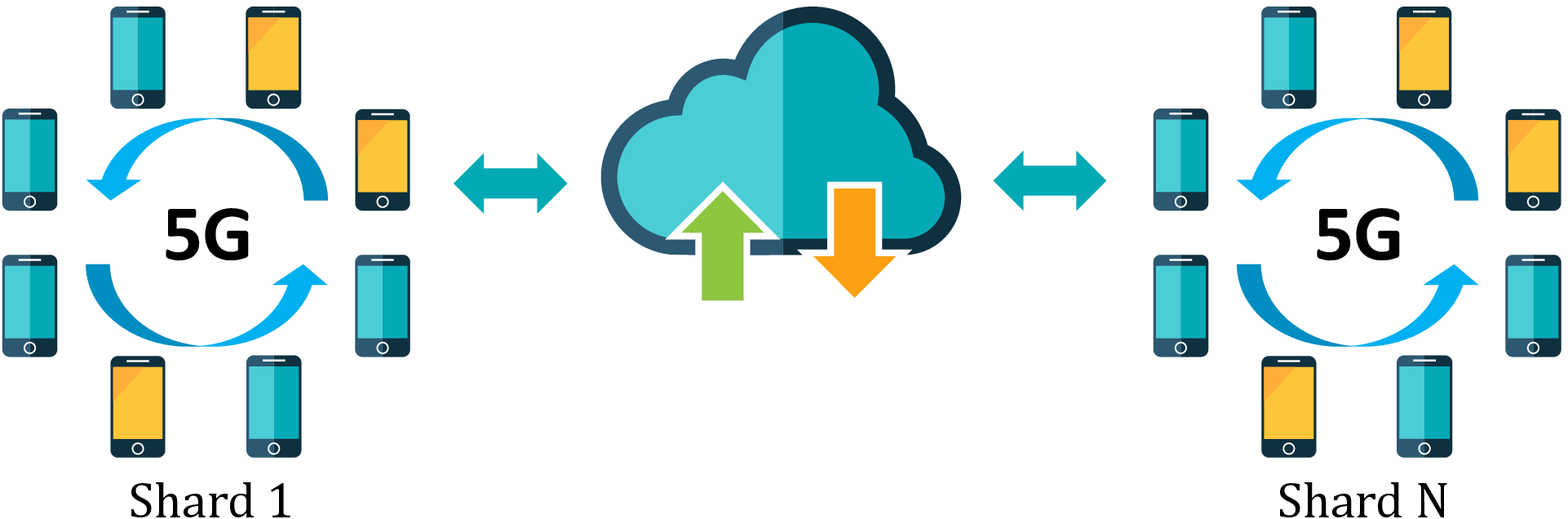}
\caption{With Sharding}
\vspace{10pt}
\end{subfigure}
\begin{subfigure}{0.485\textwidth}
\centering
\includegraphics[width=0.99\textwidth]{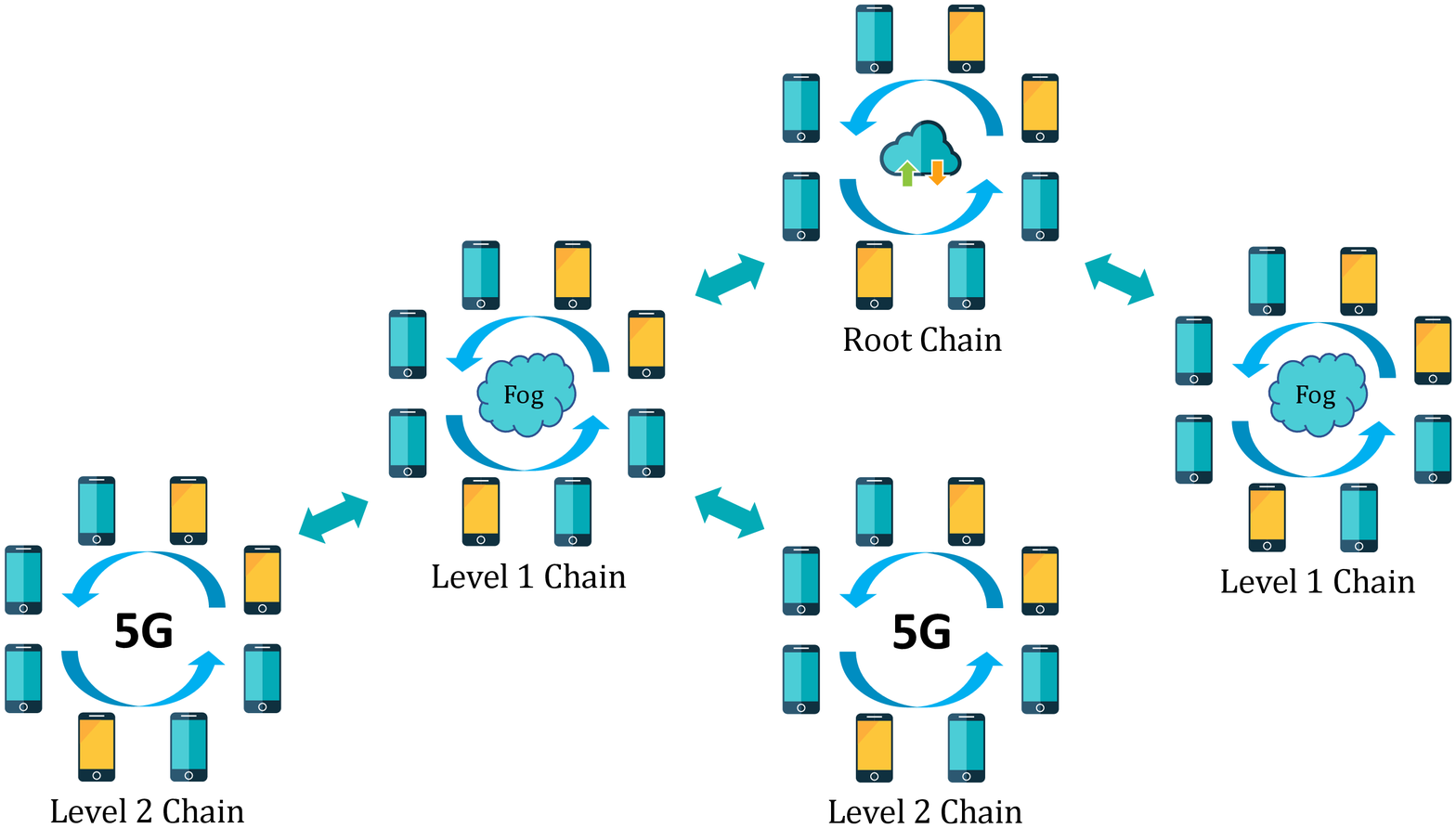}
\caption{With Hierarchical Chains}
\end{subfigure}
\caption{5G in blockchain networks}
\label{fig:5g_blockchain}
\end{figure}

\subsubsection{Overview}

In the last decade, we have witnessed the rapid proliferation of blockchain
platforms, both in public, permissionless networks and private, permissioned
setups \cite{anh_blockchain_tkde}. From the performance point of view,
blockchains are known to exhibit low transaction processing throughput, high
latency and significant energy usage
\cite{anh_blockbench,dumi_blockchain_arm_arxiv}. This low performance is, in
part, due to the costly consensus protocol, either in the form of Proof-of-Work
(PoW) or PBFT~\cite{PBFT_99}. In the context of 5G adoption, it is useful to
investigate how blockchain systems are going to be impacted.

% Use case: MNP (mobile number portability)
From the applications point of view, blockchains could help in providing trusted
services at the edge while connecting multiple mutually untrusted entities. For
example, mobile number portability (MNP) is an application where different
telecommunication companies (telcos) that do not trust each other need to work
together to offer this service to their clients \cite{mnp_blockchain}. Using
MNP, a client can keep her mobile number while switching the telecommunication
provider. For this application, the blockchain could store a unified database to
keep track of mobile numbers, client ids, and telecommunication providers.

\subsubsection{Challenges and Opportunities}

% Use case: hierarchical blockchain / side channels
With the adoption of 5G, the number of devices that can potentially connect to a
blockchain will increase significantly. Thus, traditional blockchains are
expected to exhibit even lower performance. To improve the scalability of
blockchain, researchers have looked into reorganizing the structure of the
network. There are two key approaches to do this re-organization, as depicted in
\autoref{fig:5g_blockchain}. These two approaches are (i) sharding where the
network is split into smaller partitions \cite{hung_sharding,loi_sharding} and
(ii) hierarchical chains where there is a main (root) network and many secondary
networks \cite{plasma_eth,lightning_bitcoin}.
These approaches become more relevant in the era of 5G, edge computing, and
network virtualization.

Both sharding and hierarchical networks could improve the performance of
blockchains. Shards or secondary blockchains running at the edge, in close
proximity to 5G base stations, are supposed to run much faster compared to
global networks. For example, Hyperledger Fabric 0.6 with PBFT exhibits up to
5$\times$ higher throughput in a local cluster with Gigabit Ethernet networking
compared to Google Cloud Platform distributed across 8 regions
\cite{hung_sharding}.

%%% Lin Qian & Shaofeng & Wang Wei & Xutao
\subsection{Artificial Intelligence and Federated Learning}
\label{sec:ai}

\subsubsection{Overview}

% data
With higher connection density and bandwidth, a few billion devices are expected
to be connected to the 5G network, including mobile phones, tablets, wearables,
automobiles, and drones. The increased number of interconnected devices and the
accompanying sensors will generate a tremendous amount of data on a daily basis.
At the same time, there is a surging demand for personalized services on mobile
devices to enhance user experience. For example, companies may want to provide
real-time personalized recommendations to users. The unprecedented amount of
data residing in the edge devices is the key to build personalized machine/deep
learning models for enhanced user experience. This trend presents new
opportunities as well as challenges for machine learning (ML) and deep learning
(DL).

Over the past few years, various hardware accelerators like APU (AI processing
unit), NPU (neural processing unit) and VPU (vision processing unit), have been
integrated to mobile chip platforms, including Qualcomm, HiSilicon, MediaTek,
and Samsung chipsets, to support fast inference of ML and DL models in the edge
devices. Corresponding software libraries are also developed, e.g., SNPE SDK,
Huawei HiAI SDK, NeuroPilot SDK, Android NNAPI, and TensorFlow-Lite. On the
other hand, the training is typically done on the cloud. The model is then
converted into a certain format to be deployed in edge devices. Although 5G
enables fast data transfer between mobile devices and base stations, the
edge-cloud links have a limited capacity which may not scale with the number of
connected 5G devices. Consequently, some training tasks need to be shifted from
the cloud to the edge to save the communication cost of data transfer.
Meanwhile, training in edge devices resolves the data privacy issue as the data
are not shared on the cloud.

Edge devices are expected to handle the training process in some specific
scenarios. With datasets getting larger at the edge, the training has to be
conducted locally or in the fog, rather than on the cloud, since edge-cloud
links typically have limited capacity. In this case, each device or sensor is no
longer merely a data carrier; it will also handle processing and data requests
from servers or other devices. With faster data flow between mobile devices and
5G base stations, carriers can directly transfer the required data to the client
via the edge, fog or D2D, without involving cloud servers. Even ML/DL
models can be transferred directly between devices without a server.

\subsubsection{Edge Data Properties}

Data from mobile devices has some special properties that are different from the
assumptions of traditionally centralized training. Therefore, training in edge
devices requires substantial adaptations of model design, training, and
deployment algorithms. The main properties of mobile datasets for learning in
the 5G era can be summarized as:

\begin{itemize}

\item \textbf{Highly-distributed.} The data are distributed among end devices
instead of being collected in a centralized server, where the number of end
devices would easily surpass the number of training samples per client.

\item \textbf{Unstructured.} The majority of samples in a local dataset are
expected to be in unstructured and diversified formats since raw data are
collected from various applications or sensors.

\item \textbf{Non-IID.} The local dataset is gathered from a particular client,
and thus, a great variance is expected among different local datasets. Hence,
the local dataset is not independent, identically distributed (IID) sampled
from the population distribution.

\item \textbf{Unbalanced.} The amount of training samples varies in different
clients. Moreover, sensors bias and the difference in user preferences lead to
unbalanced local datasets.
\end{itemize}

These data properties pose challenges to the traditional ML/DL training which
requires centralized structured training data \cite{singa15}. First, providing
real-time personalized services while reducing the server-side burden is highly
demanded. Second, the large number of client-side \textit{data islands}
\cite{federated_learning_19} and increased communication efficiency of 5G
connections will undoubtedly require substantial adaptation of model design,
training, and deployment. In response to these challenges, the research
community has focused on techniques for edge device architecture engineering
\cite{iandola2016squeezenet_arxiv,ma2018shufflenet,sandler2018mobilenetv2},
model compression \cite{denton2014exploiting,han2015deep_arxiv,han2015learning},
and neural architecture search
\cite{cai2018proxylessnas_arxiv,real2018regularized_arxiv,zoph2018learning}.

\subsubsection{Model Training and Deployment}

New opportunities for a wide range of research directions in AI are arising with
the adoption of 5G. In terms of model design and deployment, more client-side
models are anticipated. This is possible because (i) edge storage and compute
resources are more powerful with various system-on-chip (SoC) technologies and
(ii) there is a data-privacy practice to keep personal data locally.
Further, due to its inherent capability of adaptive modeling and long-term
planning, reinforcement learning presents potential in building interactive and
personalized models, such as interactive recommendation systems
\cite{zhao2018recommendations,zheng2018drn}.

How to build the \textit{correct} machine learning model on edge devices remains
a challenging problem. Since the computational capabilities of edge devices are
mostly limited by battery and storage space, several key factors should,
therefore, be taken into consideration for better deployment of an ML/DL model:
power consumption, storage space occupation, service latency, and model
performance. In real-world applications, these end-user perceptible requirements
and constraints should be considered in the model structure engineering and
hyper-parameter configurations during the training procedure.

Mobile edge devices also vary greatly in hardware capacity \cite{facebook_ml}.
These hardware differences require automation in the model building workflow to
satisfy the model deployment in diversified environments.
Automated machine learning and Neural Architecture Search (NAS) could provide
technical clues to tackle the challenge \cite{cai2018proxylessnas_arxiv,
tan2019mnasnet}. For example, the meta-information of datasets, local resource
profiles, and service-time constraints can be gathered to model the automation
procedure and recommend model configurations.

The properties of edge data require model training on highly biased and
unbalanced personalized data. Under such circumstances, Transfer Learning and
Meta-Learning \cite{pan2009survey} are two promising approaches to facilitate
edge training via warm start. A pre-trained global model can be either optimized
on a large public dataset or initialized via meta-learning with meta-features
collected, which is subsequently broadcasted and adapted to the local data.
Further, few-shot learning \cite{snell2017prototypical,vinyals2016matching} also
supports effective training with a minimum amount of edge data.
Knowledge distillation
\cite{hinton2015distilling_arxiv,jeong2018communication_arxiv} is another
promising technique that benefits training on edge via mutual learning, which
transfers representation learned from a high-performing teacher model to student
models that are typically smaller and more efficient for deployment.
The teacher model can also be a pre-trained global model~\cite{zhang2018deep}.
With these cutting-edge deep learning techniques, training on edge devices can
be more effective and efficient.

Other techniques, such as model compression and quantization, may benefit the
model deployment on edge devices. For instance, model compression techniques
reduce the model complexity in various ways with controlled accuracy
degradation. The quantization techniques can reduce the power consumption and
storage space cost by taking advantage of the emerging edge-optimized hardware
such as neural processing units.

\subsubsection{Federated Learning}

Increasingly, smartphones and other mobile devices become the primary computing
devices for many people. The various embedded sensors are used by popular
applications to collect an unprecedented amount of data on a daily basis. 5G
technology will undoubtedly accelerate this trend. While these data could be
used for AI model training and inference, the privacy issue should be taken into
consideration more seriously when using personal data. Moreover, directives like
GDPR \cite{gdpr} push for strict personal data processing, and require
individuals or organizations to handle data in an appropriate manner.

\begin{figure}
\centering
\includegraphics[width=0.38\textwidth]{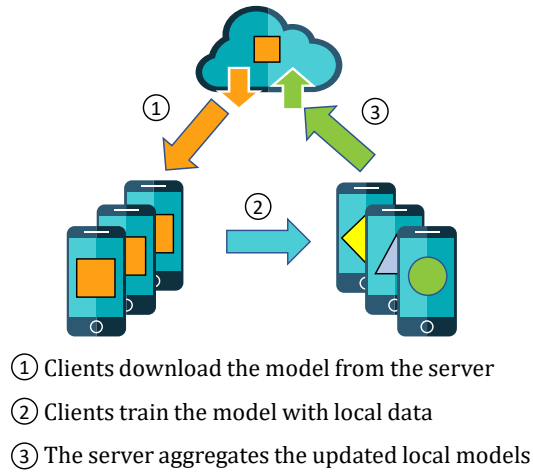}
\caption{Federated learning}
\label{fig:5g_fedlearn}
\end{figure}

To preserve data privacy, Federated Learning (FL) \cite{federated_learning_16_arxiv}
has been proposed as a collaborative training technique that keeps
the personal data residing in edge devices and constructs a shared model by
aggregating updates that are trained locally, as illustrated in
\autoref{fig:5g_fedlearn}. In federated learning, data are only accessible to
the data owner and the training process runs locally, on the mobile device. The
centralized server can only receive intermediate results such as model
updates from clients. Consequently, FL helps preserve privacy and reduce the
communication costs of dataset transfer.

Federated learning is a feasible and ideal solution for the data privacy concern
in the 5G era, where 5G and federated learning will complement each other. In
recent years, most research projects on federated learning focus on
communication efficiency and preserving privacy. The high-bandwidth and
low-latency property of 5G will improve the communication efficiency of
federated learning and compensate for the communication overhead caused by
privacy-preserving protocols. In addition, the more stable connections brought
by 5G can mitigate the dropout issues of clients during the federated learning
training. Therefore, federated learning provides a privacy-preserving solution
for learning in the 5G era, while 5G makes federated learning more practical and
robust.

Further, vertical federated learning \cite{federated_learning_19} is proposed to
tackle the data islands problem. Succinctly, vertical federated learning is a
collaborative privacy-preserving learning approach to handle scenarios where
multiple parties separately hold datasets with different attributes. For
example, when a bank and an e-commerce company decide to collaboratively train a
model without disclosing sensitive data, the datasets involved are divided into
multiple data islands and vertical federated learning comes to rescue.
In addition, smart cities are applications of great potential in the 5G era, and
different parties or platforms in a smart city will benefit from resolving the
issues related to data islands. With the 5G technology and vertical federated
learning, the world of Internet of Everything (IoE) is to be anticipated.

\begin{figure*}[tp]
\centering
\includegraphics[width=0.88\textwidth]{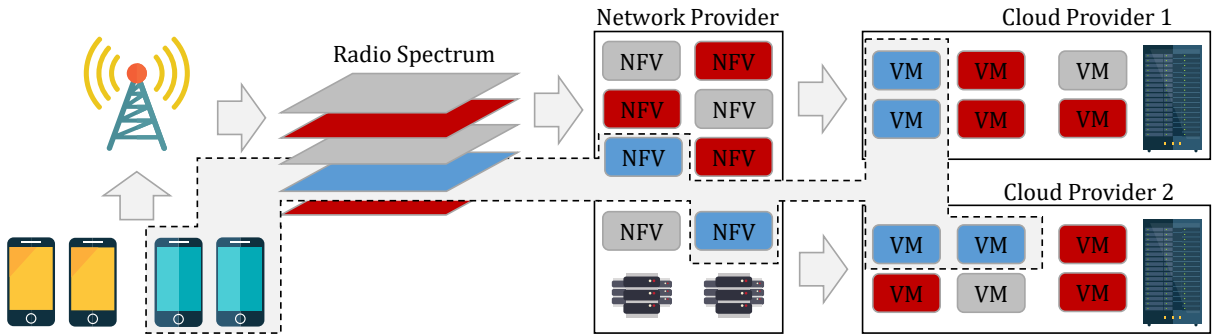}
\caption{Network slices operating at different layers need to be isolated.
Red-color sub-slices are controlled by attackers trying to learn or tamper
with data of the blue-color, honest sub-slices.}
\label{fig:isolation}
\end{figure*}

Many companies have been working on federated AI research. Google has released
TFF\footnote{\url{https://www.tensorflow.org/federated}} (TensorFlow Federated),
which is an open-source framework based on TensorFlow for machine learning and
computations on decentralized data. Meanwhile, WeBank initiated an open-source
project called FATE\footnote{\url{https://www.fedai.org}} (Federated AI
Technology Enabler). FATE provides a secure computing framework and a series of
toolkits for the federated learning ecosystem. To preserve privacy, FATE
implements secure computation protocols using homomorphic encryption and secure
multi-party computation (SMPC). However, these frameworks are still targeting
desktops, laptops, and datacenters. They have not been deployed on edge devices
mainly due to the constraints of bandwidth and computing power. But with the
adoption of 5G and the performance improvement in edge devices, we anticipate
the proliferation of federated learning platforms.

%%% Anh & Xiaokui
\subsection{Security and Privacy}
\label{sec:security}

\subsubsection{Overview}

One distinguishing feature of 5G is network slicing, which enables applications
with distinct requirements to share the same network. A generalization of
virtualization, network slicing works across all layers of the application
stack, as shown in \autoref{fig:isolation}. The radio network layer is
multiplexed through spectrum sharing. The networking layer is multiplexed at the
telco providers via SDN and NFV. Cloud resources, especially the ones near the
edge, are multiplexed via virtual machines. While virtualization has obvious
advantages in terms of better exploiting the physical infrastructure and
reducing the time to market, it poses security and privacy challenges, as we
shall further discuss.

The other features of 5G, such as improved bandwidth and latency, higher device
density and D2D communication may impact the security as well. As previously
discussed, higher device density and increased bandwidth make it easier to
conduct large scale DDoS attacks, especially using IoT devices. On the other
hand, D2D communication requires isolation and well-implemented access control
mechanisms such that data privacy is not compromised.

\subsubsection{Challenges and Opportunities}

The fact that one 5G network slice comprises multiple virtualized resources
managed by multiple providers makes it difficult to ensure isolation.
In particular, it is possible to achieve virtual machine isolation with the
secure design of hardware virtualization, but does this still hold when, for
example, slices at NFV layers are compromised? To ensure slice isolation, it
seems necessary for the layers to coordinate and agree on a cross-layer
protocol. \autoref{fig:isolation} illustrates an example where a slice consists
of several sub-slices at different layers. At each layer, the slice needs to be
isolated to ensure security and privacy.

5G is a key enabler for machine-to-machine communication. Applications based on
device location, for instance, may see new devices moving in and out of range at
high velocity. This type of ad-hoc communication with a high churn rate poses a
new challenge for device authentication. In particular, devices must establish
identities of each other before communicating, for example, by knowing the
mapping of devices to their public keys. The scale of 5G requires an identity
system that supports a large number of users and avoids a single point of trust.
Existing public key infrastructures (PKIs) are too heavyweight because they are
designed for enterprise identities. Large-scale consumer systems such as those
used for end-to-end encryption, for example, iMessage and WhatsApp, meet the
performance and scalability requirements, but they still rely on a centralized
party. To decentralize the existing identity systems, we envision a
blockchain-based solution which maintains a highly available and tamper-evident
ledger storing identity information. However, existing blockchains are severely
limited in their throughput and latency. Therefore, novel blockchain systems are
needed to meet the performance requirement of future 5G applications.

Current practices in enterprise security rely on collecting and analyzing data
both at endpoints and within the network to detect and isolate attacks. 5G
brings more endpoints and vastly faster networks. More endpoints mean a larger
attack surface, raising the probability of the network being attacked to near
certainty. Faster networks impact data collection, as it becomes unfeasible to
store, and later analyze, highly granular data over long periods of time.
Therefore, 5G demands a fundamentally new security analytics platform. We note
that existing solutions, for example
Splunk\footnote{\url{https://www.splunk.com}} and
LogRythm\footnote{\url{https://logrhythm.com}}, are inadequate for the 5G scale
since they stitch together general-purpose data analytics platforms.
The desired solution should not have been designed to target general data
management workloads, but specifically optimized for 5G workloads.

Apart from the security aspects, 5G also presents new challenges and
opportunities in terms of privacy, as the improved bandwidth and reduced latency
of 5G open up the possibility of transforming mobile devices into private
databases that could be queried in real-time. For example, consider an online
shopping service that provides recommendations to users based on their shopping
histories. With current technologies, performing such recommendations requires
the service provider to store users' shopping histories at the server-side,
which has implications for privacy. In contrast, with the help of 5G, we may
keep each user's shopping history in her local device, and let the service
provider join hands with the users to perform recommendations in a
privacy-preserving manner, e.g., by offloading to the users the part of the
recommendation task that requires access to private data.

Such a computation paradigm, however, poses a number of challenges from a
technical perspective. First, how should we manage each user's private data on
her local device, so that different service providers could access data through
a unified and efficient interface? Second, how could we enable users to make
educated decisions regarding which service provider should be allowed to access
what data item? In addition, given that each user may have a considerable amount
of heterogeneous private data stored on her local device, how could we alleviate
users' overhead in setting up access controls for a sizable number of service
providers? Third, when a service provider and a user jointly compute the result
of a certain task, the service provider may {\it infer} sensitive information
from the computation result, even if she does not have direct access to a user’s
private data. For example, based on the result of the recommendation computed
from a user’s shopping history, the service provider may infer partial
information about the items that the user purchased in the past. How should we
prevent such inference attacks without degrading the accuracy of the result
jointly computed by the user and the service provider? Addressing these issues
could lead to the development of new techniques that advance the state of the
art in privacy-preserving data analytics.

\subsection{Challenges}

\begin{figure}[tp]
\centering \includegraphics[width=0.42\textwidth]{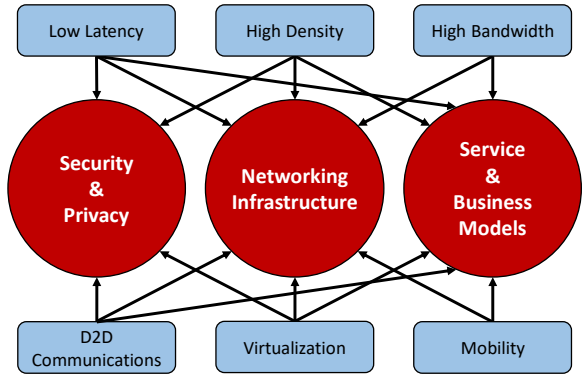}
\caption{Challenges (red) introduced by key 5G features (blue)}
\label{fig:challenges}
\end{figure}

We conclude this section by highlighting the key challenges that 5G is
introducing in areas related to data management and processing, as depicted in
\autoref{fig:challenges}.

\textbf{Security and Privacy.} Some of the key features of 5G have a significant
impact on security and privacy. First, the support for a massive number of
connections increases the area of attack and provides an ideal setup for large
DDoS attacks. Second, network slicing and end-to-end virtualization are
challenging in terms of security management in the presence of multiple service
providers. Third, D2D communication introduces security and privacy challenges
in an era when people are more concern about their private data and when strict
data protection frameworks, such as GDPR, are enforced. In this context, there
is a need for security standards and ways to ensure consensus among entities
participating in a 5G network.

\textbf{Network Infrastructure.} With its impressive bandwidth and high device
densities, 5G allows more data to be downloaded or uploaded from and to the
cloud. But this will exert a high pressure on both the (i) backhaul links from
the base stations to the rest of the operator's network infrastructure and (ii)
backbone of the Internet, including inter-cloud connections. Our networking
performance measurements among different cloud regions show that current
connections are not ready for the speeds of 5G. For example, inter-region
connections can hardly reach 100 Mbps in throughput and 10 ms in delay, in the
best case, while 5G specifications require at least 10 Gpbs and 1 ms throughput
and latency, respectively (\cref{sec:5g_specs}). We assert that there is a need
for both (i) better backbone connectivity and (ii) smart edge-fog-cloud data
offloading strategy to cope with the demands in services, data movement, and
data processing.

\textbf{Service and Business Models.} With the explosion of the number of IoT
and mobile devices connected to the Internet through 5G, there is a need for new
service delivery and business models. 5G is considered the ideal network for
connecting IoT devices, but creating a separate subscription for each device may
be inconvenient for the user. On the other hand, operators will need to invest
significantly to improve their network infrastructure~\cite{5ginfra} and to be
able to deliver quality services at the edge. In addition to the pressure of
high 5G bandwidth on the backhaul network, the high mobility specific to mobile
devices introduces new challenges in service delivery and accountancy, in the
context of virtualization and edge computing.

\section{Use Cases and Challenges}
\label{sec:use_cases}

% wangwei, linqian, zhonghua, janiceng, dumi

While the previous section analyzed the impact of 5G on areas related to data
management and processing, this section presents 5G use cases with a focus on
analyzing challenges and identifying research opportunities.

\subsection{Healthcare}
\label{sec:healthcare}

The healthcare industry is rapidly expanding, mainly due to the advancements in
machine learning which are applied to the medical domain \cite{Lee2017}. In a
recent study, Deloitte estimates that the healthcare market will grow to 10
trillion US dollars by 2022 \cite{deloitte_healthcare_19}. With the adoption of
5G, new smart healthcare use cases are taking shape, such as telemedicine,
telesurgery, and smart medical devices.

% Telemedicine
5G will be the foundation of telemedicine in countries where wired
infrastructure is not well developed. 5G mobile services will enable more
effective delivery of remote diagnosis and support for paramedics. This allows
for a new and seamless way of delivering cost-effective and direct-to-consumer
healthcare as it is no longer limited to traditional face-to-face consultations
in healthcare facilities. In order to have connected care and telemedicine, 5G
is needed to guarantee low latency and high-quality video streaming.

% Telesurgery
Telesurgery can also benefit from the low latency and high bandwidth of 5G.
Telesurgery allows surgeons to execute real-time surgery, even when they are not
physically in the same location, using a remote control to carry out the
surgery. Although 4G is sufficient for real-time video transmission under ideal
conditions, its relatively high latency renders it unusable for telesurgery.
It remains to be studied if 5G, with its improved latency and increased
bandwidth, is able to meet the requirements of telesurgery.

% Smart medical devices
One of the main reasons patients with chronic diseases visit the hospital is the
lack of medical equipment at home to measure and monitor vital body signs.
5G will alleviate the burden of hospital checks by transferring this
functionality to the community (e.g., to local clinics and homes). Devices that
are community-deployable should be equipped with vital signs sensing, biomarker
sensing, video analytics, a chatbot and an AI-enabled intervention mechanism
(e.g., a model that can predict disease progression
\cite{Lee2017,Kaiping_DPM_17}). All these features are more feasible in the 5G
era.

% IoMT
Massive Internet of Medical Things (IoMT) market is predicted to grow from 8 to
33 million shipments in the period 2016-2021 \cite{medical_wearables_17}.
IoMTs are clinical wearables consisting of low-power medical monitoring devices
that allow for tracking a patient's status. Such an integrative device receives
information from various sensors and sends pre-processed data to healthcare
providers who may adjust the medication doses or change the behavioral therapy.

% Challenges
We assert that the security and privacy challenges in the era of 5G pertain to
the field of healthcare and IoMT, as well. With a series of recent security
breaches in medical data management systems \cite{uk_data_breach,
sg_data_breach}, security and privacy are one of the biggest concerns in the
digitalization of healthcare. Moreover, strict personal data processing
directives, such as GDPR \cite{gdpr}, require special attention. It remains to
be studied if 5G virtualization could address these concerns.

\subsection{Smart City}
\label{sec:smart_city}

% What is smart city
The key features of 5G, such as high speed, massive connections, and
virtualization, will enable the development of smart cities. A \textit{smart
city} is a sustainable city that utilizes smart solutions to improve the
infrastructure and provide better services to the community
\cite{albino2015smart}. Among smart city solutions, we mention correlated
traffic systems, public safety, security, and surveillance. A key objective of a
smart city is to provide cohesion among the variety of deployed systems.

% Smart city use cases
Below, we enumerate a few smart city applications that may be enhanced by 5G
technologies. Firstly, \textit{smart homes} can be implemented with many
interconnected devices and with fast Internet access which is needed for
security monitoring. Secondly, \textit{smart education} could be enabled by
stable connectivity and high bandwidth. Students will be able to access a
massive number of online courses and even participate remotely in real-time
classes. Thirdly, \textit{smart safety and surveillance} could be enabled by
reliable connections and the integration of real-time video observation from
various locations. This allows real-time emergency response and surveillance of
traffic conditions, accidents, banks and ATMs, stores, roads, among others.
Lastly, \textit{smart power} could be implemented using \textit{smart grid}
technology \cite{smart_grid} consisting of smart meters, sensors, and data
management systems. A smart power solution reduces energy and fuel consumption,
while identifying power outages in real-time.

% Current challenges in smart city - reliability
Currently, smart cities are not efficiently implemented due to a lack of
powerful connectivity \cite{Rao2018}. Low-latency, stable connectivity is
required anywhere and anytime within a smart city. It is estimated that the
reliability of the network in a smart city should be higher than
99.9999\%~\cite{Rao2018}. Moreover, the network infrastructure of a smart city
must be able to support an immense amount of IoT devices. 5G suits the
requirements of smart city connectivity, with its low latency of 1 ms, and high
device density of up to one million devices per square kilometer.

% Challenge - energy efficiency
Another challenge in smart cities is ensuring the energy efficiency of
monitoring solutions \cite{smart_city_monitoring}. This is challenging in the
context of maximizing the life of battery-operated sensors and requires a smart
deployment of devices, as well as algorithms to compute an optimal
communication-to-computation ratio per device. Nonetheless, the energy
efficiency of 5G terminals could improve the battery life of remote monitoring
devices.

% Challenges introduced by 5G
While the benefits of 5G in smart cities are obvious, some challenges need to be
addressed. First, with the interconnection of vital city infrastructure, there
is a high security risk in case attackers manage to capture critical nodes.
Network slicing is a partial solution to this, where different smart city
applications are isolated. However, we discussed in \autoref{sec:security} that
5G virtualization presents some security risks that need to be addressed.
Second, the high volume of data from surveillance and monitoring systems will
exert high pressure on the network infrastructure connecting 5G base stations
with central facilities. A solution to this is the use of edge and fog computing
where partial processing with the discarding of fruitless data can be done
closer to the source of data.

\subsection{Automotive}
\label{sec:automotive}

The automotive industry will be significantly impacted by 5G, as it opens up the
potential for vehicles to be connected to roadside infrastructure, pedestrians,
and other vehicles. Currently, autonomous vehicles are not fully supported by
the IT infrastructure due to the lack of mobile antennas and sensors, which does
not allow for efficient and stable communications \cite{choi2016millimeter}.

% 4G not enough
4G technology is unable to reach the handling, processing, and analyzing
standards needed by autonomous vehicles \cite{choi2016millimeter,
keysight_autonomousvehicle,5g_smart_car}. In order for autonomous cars, also
known as smart cars or self-driving cars, to be well-implemented, the time to
transmit and process sensor data needs to match at least the speed of human
reflexes \cite{5g_car_bbc}.

% 4G not enough
Existing 4G infrastructure, including the mobile antennas on buildings, is not
sufficient for autonomous cars~\cite{5g_car_bbc}. There is a need for
significant amounts of antennas located a few hundred meters apart to enable
stable car-to-car communications~\cite{5g_car_bbc}. 5G, with its D2D technology,
could help in alleviating this issue. In addition, D2D helps with sensor fusion
such that cars can have a better view of the traffic and road condition beyond
their line of sight.

% V2V
Wireless communication enables vehicles to share, among them or with other
participants, information about road and traffic conditions. For example,
Cellular Vehicle-to-Everything (C-V2X) protocol comprises of multiple
communication methods, such as Vehicle-to-Vehicle (V2V),
Vehicle-to-Infrastructure (V2I), Vehicle-to-Network (V2N), and more
\cite{v2x_automotivedomain}. 5G's low latency will allow for V2V and vehicle
platooning, where vehicles communicate directly to share warnings and real-time
road conditions. V2I enables communication between vehicles and roadside
infrastructure components, such as traffic signs, traffic lights, and pedestrian
crossings. The predicted reliability of 5G at 99.9999\% will allow for V2N to
run smoothly as it can share real-time traffic information with the wireless
network infrastructure.  These smart vehicle technologies can anticipate
potential risks or help in planning an optimal route given real-time traffic
conditions. Moreover, these technologies are predicted to improve safety and
reduce deaths, since 90\% of fatal car accidents are due to human error
\cite{ieee_automotivedomain}.

% Challenges introduced by 5G
While 5G is seen as the natural choice for wireless communications in autonomous
cars, there are some challenges that need to be addressed. Firstly, critical
decisions must be taken by the autonomous car based on its own processing, such
that the reaction time is kept below 2 ms \cite{5g_car_bbc}.
Even if 5G has a theoretical latency of 1 ms, this is the best-case latency to
the base station. If multiple hops are needed to get the required data, the
latency may increase above 2 ms. For example, our measurements on a real 5G
network show an RTT of 6 ms to a sever that is a few hops away. Secondly, there
is the challenge of trust and authenticity in the messages received by a vehicle
from other entities. In the context of security issues in 5G environments,
discussed in \autoref{sec:security}, there is an imperative need to evaluate
their impact on critical systems, such as autonomous vehicles.

\subsection{Smart Drones}
\label{sec:smart_drone}

% Use cases of drones: base stations
The flexibility in the deployment of unmanned aerial vehicles (UAV), also known
as drones, has enabled a series of use cases such as the spread of the Internet
in remote areas, public safety communications, disaster recovery, flood area
detection, and special deliveries. The use of multiple drones, or drone swarm,
allows for the spread of the Internet to areas that lack reliable connectivity.
In this use case, multiple drones fly autonomously in close proximity to build a
wireless network with no gaps in signal distribution to the ground
\cite{5gDrones}.

% Opportunity - large radio spectrum and mmWave
The deployment of UAV base stations (e.g., drone base stations) \cite{uavbs},
can be accelerated by 5G, especially with the usage of the mmWave technology and
a massive number of connections. Currently, the limited radio frequency spectrum
below 6 GHz is not capable of supporting smart drones and UAVs. With the use of
a larger spectrum, between 28 and 95 GHz (\cref{sec:5g_specs}), 5G enables
effective communication between drones and ground users. More specifically, 5G
will enable wireless mobile broadband with low latency and high connection
density.

% Opportunity - energy efficiency
Another feature of 5G, namely energy efficiency, could have an impact on UAV
base stations. These UAVs are battery-operated and need to exhibit satisfactory
operating time to enable reliable connectivity \cite{5gDrones}. While
alternative sources of energy, such as solar panels, can be used to enhance
battery life, the energy efficiency of 5G is a complementary feature that can
extend operating time.

% Use case 2: disaster recovery
5G-connected drones can aid in emergencies, where drones communicate and share
real-time information with operators on the ground. This increases the success
of the search and rescue missions and allows relief teams to dispatch rescue
teams. For example, this use case can quickly estimate debris levels and
distribute resources efficiently.

% Challenges
The adoption of smart drones and 5G will expose challenges related to security,
privacy, and public safety \cite{drones_smart_city}. In addition to DDoS
cyber-attacks, such as those launched using IoT devices (\cref{sec:edge}),
drones could be used to conduct physical attacks on the population of a smart
city. In the context of UAV base stations, there needs to be a clear separation
between the Internet providing service and the UAV control plane. Network
slicing and virtualization could help in addressing some of the security
challenges of smart drones.

% \subsection{Remote Desktop Computing and Virtual Reality}
\subsection{Virtual and Augmented Reality}
\label{sec:vr}

% AR/VR is requested by users and 5G can enable it
In a study by IBM's Institute for Business Value, prospective users of 5G are
looking forward to entertainment applications based on VR and AR
technologies~\cite{ibm_ibv_study}. The potential use cases of these interactive
and immersive technologies are wide and varied, but the platform behind these
revolutionary technologies is the same: a combination of cloud, edge, and 5G
connectivity~\cite{VR_AR_cloud}. Currently, the challenges faced by VR, AR, and
mixed reality are mainly related to the lack of mobility and bad user experience
in terms of lag and low video quality. Under 5G, the distributed edge computing
will be the main technology to tackle those issues. With the high bandwidth and
low latency of 5G, cloud and edge computing can deliver the high-resolution
content to the VR glasses, while enabling computation offloading from the VR
glasses to the edge cloudlets or directly to the cloud. In this way, content
delivery will be faster, enabling smooth VR/AR experiences.

% gaming, live streaming, reality show
As part of the AR/VR entertainment, users are interested in interactive gaming
and immersive streaming of sports, e-sports, and reality
shows~\cite{ibm_ibv_study}. With 5G, gaming experience improves due to low
latency connections to gaming servers. At the same time, the high bandwidth of
5G allows graphics processing to happen on powerful servers on the cloud or at
the edge, while the high definition video is transferred back to the gamer's
device. A study by Intel and Ovum predicts that 5G will enable revenues of
almost 50 billion US dollars by 2028 in the AR/VR gaming
industry~\cite{intel_ovum_study}.

% remote work
Beyond entertainment, 5G will connect the front-end and back-end workers in big
organizations. Front-end workers are always the first to interact with a
potential or existing customer or make a product demo for the company. It is
often critical for big organizations to connect the customer, the front-end
worker, and the leader, across geographic boundaries. With 5G, communication
tools will support real-time feedback which allows distributed workers to
overcome the communication delay and respond to customer needs timely. This is
very useful, especially in a fast-paced environment.

Moreover, we envision that 5G will enable better experiences in working
remotely. 5G latency is much lower compared to the refresh rate of ordinary
displays, which is 60 Hz (or 17 ms). In such cases, the terminals connected to
5G do not need to be fat clients running an entire operating system: they can be
low-power devices equipped with a simple browser. The high bandwidth of 5G
allows UHD graphics streaming, where the 3D graphics engine runs on a remote
server, for example in a cloudlet.

Huawei has started its cloud desktop service for both enterprise clients
\cite{hw_enterprise_cloud_desktop} and individual clients \cite{hw_cloud_pc}. In
a 5G environment, this cloud desktop service will support image quality of up to
4K. In this way, a mobile device connected to a 5G network can serve as a
portable workstation or as a mobile game console. Nvidia introduced the RTX
server for cloud-based GPU computing \cite{nvidia_rtx_server}. At the moment,
these services do not deliver excellent user experience, especially for mobile
users, because of high network latency. However, we expect an improvement with
the adoption of 5G.

\subsection{E-commerce and Fintech}

% immersive e-commerce experience
E-commerce, such as online shopping, will be further disrupted by the 5G
technologies. High-quality video streaming and real-time information feed will
not only provide an immersive shopping experience with fast and personalized
recommendations but also enable dynamic mix-and-match
choices~\cite{forbes_ecommerce}. For example, a piece of furniture may be viewed
from the perspective of a real home environment. While 4G popularized online
shopping, 5G is likely to take it a few steps further with augmented reality,
fast fact-checking, recommendations, and overall experience.

% 5G in fintech
5G, together with other technologies such as AI, IoT, and blockchain, will
disrupt the e-commerce and financial industries in the near future. E-commerce
and fintech companies will challenge traditional or legacy banks with their
offer of better online customer experience through 5G and AI. The demand for
seamless digital banking experience will transform banking services. Online
transfers, payments, and purchasing of banking products will be the norm, and
customer loyalty will become weak due to higher expectations and ease of moving
the funds around.

% 5G and IoT in banking
5G, edge computing, and IoT are going to improve both the traditional and
digital banking experiences. First, 5G with its low latency and network
virtualization technology can enable safe yet flexible placement of automated
teller machines (ATM) and point of sale (POS) in remote areas in smart cities or
the countryside. Second, IoT devices, such as smart watches and smart wallets,
will use 5G for faster and safer banking.

% 5G and AI in trading
AI has been used to improve the profit in trading by examining historical
records, relevant news and information, and model performance~\cite{ai_trading}.
Trading bots are becoming more intelligent, being able to maximize profits and
make smaller loses. While human traders are still dominant, AI algorithms are
being increasingly deployed by trading companies. With 5G, real-time
collaborative trading, either among humans or between humans and AI, becomes
feasible due to low networking latencies.

\begin{figure}[tp]
\centering
\includegraphics[width=0.485\textwidth]{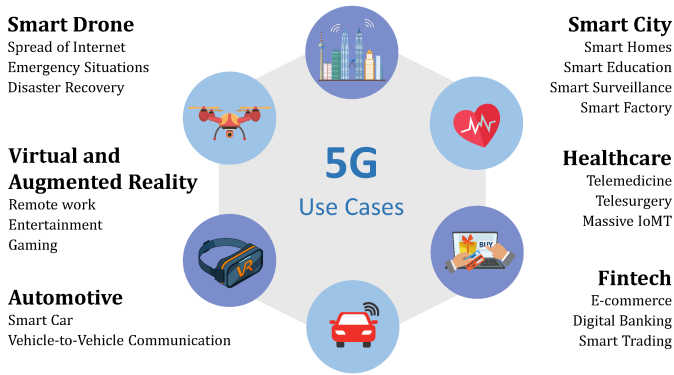}
\caption{5G Use cases overview}
\label{fig:usecases}
\end{figure}

\subsection{Summary}

Based on an extensive literature review, we have presented in this section some
key use cases that are going to be enabled by 5G, as illustrated in
\autoref{fig:usecases}. We summarize our presentation by highlighting the trends
and challenges we foresee in the event of 5G adoption.

\textbf{Efficient Healthcare.} The healthcare sector has a huge market size
which is going to increase with the population's aging all over the world. 5G,
together with Machine Learning and IoMT, is going to enable more efficient and
affordable healthcare, even in under-developed countries. However, the challenge
in remote healthcare is represented by the security and privacy of patients'
data.

\textbf{Smart City.} Smart cities, including smart cars, smart drones, and smart
grids, are going to benefit from 5G, as it reduces latency, enables massive IoT,
and offers highly-reliable connectivity. Again, the main challenge is
represented by the security risks associated with the adoption of these
technologies. It remains to be investigated if virtualization and network
slicing in 5G are going to alleviate the security risks or introduce new issues.

\textbf{Virtual and Augmented Reality.} Virtual and augmented reality is a
sector with huge business potential that spans both entertainment and
work-related activities. With its increased bandwidth and low latency, 5G will
create an immersive experience in movie and live streaming, gaming, reality
shows, among others. On the other hand, 5G will increase the productivity of
businesses that use remote desktop computing.

\section{Conclusions}
\label{sec:conclusion}

With 5G on the verge of being adopted as the next mobile network, it is
necessary to analyze its impact on the landscape of computing and data
management. In this paper, we have analyzed the broad impact of 5G on both
traditional and emerging technologies and shared our view on future research
challenges and opportunities. We hope this review serves as a basis for further
study and development of relevant technologies. 5G will make the world even more
densely and closely-connected, and will present us with vast amounts of
possibilities and opportunities to overcome the challenges ahead of us.

\vspace{16pt}
\noindent \textbf{Acknowledgement}: We thank Dan Banica for helping us with 5G
measurements in Bucharest. This research is supported by Singapore Ministry of
Education Academic Research Fund Tier 3 under MOE’s official grant number
MOE2017-T3-1-007. Tien Tuan Anh Dinh is supported by Singapore University of
Technology and Design's startup grant SRG-ISTD-2019-144.

\vspace{20pt}
\appendices
\section{}
\label{sec:appendix}

\subsection{5G Measurements}
\label{sec:apdx_5g}

\begin{figure}[t]
\centering
\includegraphics[width=0.45\textwidth]{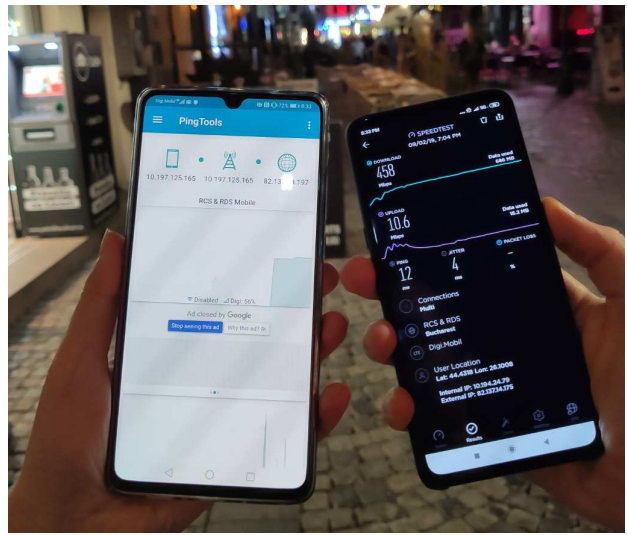}
\caption{Measurements setup}
\label{fig:meas}
\end{figure}

In this section, we extend the measurements presented in
\autoref{sec:5g_deployments}.
These measurements were run in an existing 5G deployment operated by RCS/RDS in
Bucharest, Romania. The base stations employed by this 5G network are produced
by Ericsson \cite{5g_digi_ericsson}, while the 5G smartphones are represented by
(i) Xiaomi Mi Mix3 5G (\textbf{Mix3}) \cite{mix3} with a Qualcomm X50 5G modem
and (ii) Huawei Mate 20 X (\textbf{MateX}) \cite{matex} with Huawei's 5G modem.
We use Ookla's Speedtest Android application to run our measurements, as shown
in \autoref{fig:meas}. We selected three test servers hosted by major telcos in
Romania. The results, summarized in \autoref{tab:5g_measurements}, compare
download and upload throughput, latency, and jitter of 5G and 4G networks.

First, the results show a significant difference between the download throughput
of 5G and 4G. The difference in peak download throughput is almost 5$\times$,
with 458 Mbps for 5G and 86.5 Mbps for 4G. At the other extreme, the difference
between the lowest download throughput is 35$\times$, with 244 Mbps for 5G and
6.9 Mbps for 4G. We observe a high variability in 4G download throughput when
the Mix3 smartphone is used: the highest and lowest values are 86.5 Mbps and 6.9
Mbps, respectively. We attribute this to different network conditions during
measurement collection, such as different base stations to which the device was
connected, device location, network congestion, among others. While we tried to
use the same location for all tests, we were not always connected to the same
base station, based on the gateway's IP displayed on our phones.

The second observation is that the upload throughput is similar on 5G and 4G,
especially when the Mix3 smartphone is used, with values in the range 3.4-23.9
Mbps. With MateX, the maximum 5G upload throughput is 3.5$\times$ lower compared
to the maximum 4G upload throughput. This suggests there are issues with the 5G
modem or its configuration on the phone, besides the asymmetrical bandwidth
provided by the telcos. Nonetheless, 5G upload throughput is well below the 10
Gbps peak provided by the specifications \cite{min_int2020}.

The third observation is that 5G latency is relatively high, being in the range
of 6-41 ms in our measurements. With Mix3, the measured 5G latency (11-41 ms) is
comparable to the measured 4G latency (13-48 ms). In contrast, MateX exposes
lower latency and jitter. While the measured latency values represent the RTT,
they are still far from the 2 ms RTT based on the specifications
\cite{min_int2020}.

% Server #1 - RCS/RDS
% Server #2 - UPC
% Server #3 - Orange
\begin{table}[tp]
\centering
\caption{Measurements on 5G and 4G Networks}
\label{tab:5g_measurements}
\resizebox{.49\textwidth}{!}{
\begin{tabular}{|c|c|r|r|r|r|r|r|r|r|}
\hline
\multirow{3}{*}{\rotatebox[origin=c]{90}{}} & Device &
\multicolumn{4}{c|}{\textbf{Xiaomi Mi Mix3 5G}} &
\multicolumn{4}{c|}{\textbf{Huawei Mate 20 X}} \\
\cline{2-10}
& \multirow{2}{*}{Server} & Download & Upload & Latency & Jitter & Download
& Upload & Latency & Jitter \\
& & \multicolumn{1}{c|}{[Mbps]} & \multicolumn{1}{c|}{[Mbps]} &
\multicolumn{1}{c|}{[ms]} & \multicolumn{1}{c|}{[ms]} &
\multicolumn{1}{c|}{[Mbps]} & \multicolumn{1}{c|}{[Mbps]} &
\multicolumn{1}{c|}{[ms]} & \multicolumn{1}{c|}{[ms]} \\
\hline
\hline
\multirow{9}{*}{\textbf{5G}} & \#1 & 458 & 10.6 & 12 & 4 & 458 & 8.5 & 6 &
1\\
& \#1 & 369 & 11.8 & 18 & 125 & 434 & 6.8 & 7 & 2 \\
& \#1 & 351 & 13.3 & 12 & 5 & 413 & 7.5 & 8 & 3 \\
\cline{2-10} & \#2 & 413 & 12.9 & 12 & 11 & 445 & 8.1 & 7 & 4 \\
& \#2 & 357 & 9.7 & 11 & 20 & 375 & 9 & 7 & 1 \\
& \#2 & 258 & 11.5 & 12 & 3 & 343 & 6.1 & 9 & 2 \\
\cline{2-10} & \#3 & 309 & 8.5 & 14 & 145 & 415 & 4.6 & 8 & 3 \\
& \#3 & 303 & 11.8 & 41 & 1 & 389 & 8 & 9 & 2 \\
& \#3 & 244 & 20.6 & 10 & 6 & 347 & 7 & 8 & 1 \\
\hline
\multirow{9}{*}{\textbf{4G}} & \#1 & 86.5 & 11.8 & 32 & 1 & 37 & 28.6 & 33 &
2 \\
& \#1 & 18.1 & 11.2 & 14 & 1 & 26 & 27.6 & 33 & 3 \\
& \#1 & 6.9 & 13.9 & 13 & 3 & 9.9 & 28.3 & 31 & 8 \\
\cline{2-10} & \#2 & 86.3 & 11.5 & 48 & 1 & 45.7 & 27.2 & 51 & 3 \\
& \#2 & 17.0 & 3.4 & 15 & 4 & 38.1 & 29.2 & 54 & 5 \\
& \#2 & 9.3 & 23.9 & 17 & 2 & 33.6 & 23.7 & 44 & 5 \\
\cline{2-10} & \#3 & 29.7 & 10.5 & 15 & 3 & 42.2 & 28.4 & 47 & 5 \\
& \#3 & 8.4 & 13.4 & 16 & 3 & 40.9 & 29.6 & 44 & 3 \\
& \#3 & 7.8 & 18.1 & 16 & 6 & 38 & 23.3 & 49 & 0 \\
\hline
\end{tabular}
}
\end{table}

We attribute these results to at least two factors, as discussed in
\autoref{sec:5g_deployments}. First, current 5G deployments run in the
Not-Stand-Alone (NSA) mode \cite{5G_nsa_ericsson}, together with 4G networks
and, thus, exhibit lower performance compared to the specifications of
Stand-Alone 5G. Second, the test servers are more than one hop away from the
phone. In our measurements, two of the servers are at 9 and 13 hops away from
the phone, respectively. Usually, the backhaul links are based on fiber or
microwave \cite{5gbackhaul} and have high capacity, but other network parameters
and conditions (e.g., routing, congestion) may affect the performance. The type
and configuration of backhaul links could explain the low upload throughput of
5G because operators tend to prioritize the download throughput which is more
perceptible to the end-user.

\end{document}